\documentclass[reprint,groupedaddress,showpacs,preprintnumbers,nofootinbib,nobibnotes,amsmath,amssymb,aps,prd,floatfix,superscriptaddress]{revtex4-1}

\usepackage[colorlinks=false]{hyperref}
\usepackage{dcolumn}% Align table columns on decimal point
\usepackage{bm}% bold math
\usepackage[normalem]{ulem}

\usepackage{graphicx,color,xcolor}
\usepackage{enumitem}
\usepackage[utf8]{inputenc}
\usepackage{mathrsfs}

\begin{document}

\title{A new temperature dependent hyperonic equation of state: application to rotating neutron star models and $I$-$Q$-relations.}

\author{Miguel Marques }\email{miguel.marques@obspm.fr}
\affiliation{LUTH, Observatoire de Paris, PSL Research University,
  CNRS, Universit\'e Paris Diderot, Sorbonne Paris Cit\'e, 5 place
  Jules Janssen, 92195 Meudon }%
\author{Micaela Oertel}\email{micaela.oertel@obspm.fr}
\affiliation{LUTH, Observatoire de Paris, PSL Research University,
  CNRS, Universit\'e Paris Diderot, Sorbonne Paris Cit\'e, 5 place
  Jules Janssen, 92195 Meudon }%
\author{Matthias Hempel}\email{matthias.hempel@unibas.ch}
\affiliation{Universität Basel, Departement Physik, Klingelbergstr. 82, 4056 Basel, Switzerland}
\author{J\'er\^ome Novak}\email{jerome.novak@obspm.fr}
\affiliation{LUTH, Observatoire de Paris, PSL Research University,
  CNRS, Universit\'e Paris Diderot, Sorbonne Paris Cit\'e, 5 place
  Jules Janssen, 92195 Meudon }%

\date{\today}% It is always \today, today,
             %  but any date may be explicitly specified

\begin{abstract}
  In this work we present a newly constructed equation of state (EoS)
  --applicable to stellar core collapse and neutron star mergers--,
  including the entire baryon octet. Our EoS is compatible with the
  main constraints from nuclear physics and, in particular, with a
  maximum mass for cold $\beta$-equilibrated neutron stars of
  $2 M_\odot$ in agreement with recent observations. As an application
  of our new EoS, we compute numerical stationary models for rapidly
  (rigidly) rotating hot neutron stars. We consider maximum masses of
  hot stars, such as proto-neutron stars or hypermassive neutron stars
  in the post-merger phase of binary neutron star coalescence. The
  universality of $I$-$Q$-relations at nonzero temperature for fast
  rotating models, comparing a purely nuclear EoS with its
  counterparts containing $\Lambda$-hyperons or the entire baryon
  octet, respectively, is discussed, too. We find that the $I$-$Q$
  universality is broken in our models when thermal effects become important,
  independent on the presence of entropy gradients.  Thus, the
  use of $I$-$Q$ relations for the analysis of proto neutron stars or
  merger remnant data, including gravitational wave signals from the
  last stages of binary neutron star mergers, should be regarded with
  care.
\end{abstract}

% PACS,

%\keywords{Suggested keywords}%Use showkeys class option if keyword
                              %display desired
\maketitle

%\tableofcontents

%%%%%%%%%%%%%%%%%%%%%%%%%
%			%
%     Section 1		%
%			%
%%%%%%%%%%%%%%%%%%%%%%%%%

\section{Introduction}
Neutron stars are among the most extreme objects in the universe. They
represent unique laboratories for probing strongly interacting matter
at ultra high densities --exceeding that in atomic nuclei--, as well
as gravity for strong fields.  They are formed in a core-collapse
supernovae (CCSN) and cool down mainly by neutrino emission to form a
catalyzed cold neutron star on a timescale of several minutes.  Thus,
in the early post bounce phase, as proto-neutron stars (PNS), they do
not contain only ultra-dense matter, but they are hot objects, too,
reaching temperatures of the order $\sim 50$
MeV~\cite{Burrows1986,Keil1995,Pons1999}. In addition, matter in a PNS
is not transparent to neutrinos, being thus lepton rich. Temperature
and lepton content are important ingredients to describe the physics
of PNSs, be it matter composition and stability of the PNS against
collapse to a black hole~\cite{Keil1995, Baumgarte1996, Pons2001,
  ishizuka_08, sumiyoshi_09, Nakazato10a, nakazato_12, hempel12,
  Peres_13, steiner13, char2015} or dynamical properties such as
frequencies and damping times of quasi-normal modes and consequently
the emitted gravitational wave signal~\cite{FerrariGW}.

In the post-merger phase of a binary neutron star coalescence, a
rapidly rotating neutron star could be formed which temporarily
resists to a black hole collapse~\cite{sekiguchi11}, even if its mass
exceeds the maximum mass of a cold non-rotating neutron star. Within
these merger remnants, temperatures of the same order as for CCSN and
PNSs are reached. Both, PNSs and merger remnants can rotate at rather
high frequencies, with potentially a differential rotation profile.

The temperatures of 50-100 MeV, reached in these astrophysical
environments are such that thermal effects on the EoS become
important.  They have in particular a non-negligible effect on the
composition, favoring the production of non-nucleonic degrees of
freedom such as hyperons, nuclear resonances, or mesons. Even a
transition to the quark-gluon plasma could take place.  The impact of
these additional particles on the evolution of PNSs has a long
history, see e.g.~\cite{Prakash1996} for an early review.  Most
models employ EoS for homogeneous matter, neglecting inhomogeneous
matter in the outer layers and the formation of a crust, see
e.g.~\cite{Pons:2000xf, Pons:2000iy, nicotra_06, Menezes:2007hp,
  Dexheimer:2008ax, yasutake09, Bombaci:2011mx, Burgio11,chen_12}.

Currently, only a few EoSs are available covering in a consistent way
the whole necessary domain in temperature $T$, baryon number density
$n_B$ and electron fraction, $Y_e = n_e/n_B$ where $n_e$ is the
electron number density. We will call them general purpose EoS. In the
last years, a series of new EoS models has been developed, see
e.g.~\cite{botvina08, Typel2009, Hempel09, GShen1, GShen2,
  Raduta:2010ym,steiner13,hempel12,furusawa13_eos,Togashi2017,Furusawa2017}, focused mainly on
the treatment of the inhomogeneous part and correct nuclear
abundances, and/or nuclear interactions at high densities.  Triggered
by investigations of stellar black hole formation, some effort has
recently been devoted to extend the existing purely nuclear models to
include non-nucleonic degrees of freedom --hyperons, pions or quarks--
at high densities and temperatures, too, see e.g.~\cite{ishizuka_08,
  nakazato08, sagert_09, Shen:2011qu, Oertel:2012qd,
  Gulminelli:2013qr, Banik:2014qja}. The latter are very important for
the description of PNSs and merger remnants in view of the high
densities combined with high temperatures which are attained within
these objects. However, up to now none of these extended models is
really satisfactory, since either not compatible with constraints from
nuclear physics or neutron star masses~\cite{Demorest2010,
  Antoniadis2013, Fonseca2016}, or containing only a limited selection
of additional degrees of freedom, typically $\Lambda$-hyperons.  Here,
we will present for the first time an EoS taking into account the
entire baryon octet and being well compatible with the main present
constraints.

As an application of our new EoS, we will compute stationary models of
(rotating) hot stars and study the influence of hyperons on PNS and
merger remnant properties.  Most studies of PNS evolution are based on
sequences of quasi-equilibrium models\footnote{However, see
  Ref.~\cite{Suwa2013} for a first dynamical study.}, an assumption
which is well justified in view of the hydrodynamic timescale
($\lesssim 10^{-3} $~s) being much smaller than the timescale on which
thermodynamic properties are modified considerably ($\approx 1$~s),
see e.g.~\cite{Burrows1986, Pons2001, Villain2004}. Although merger
remnants cannot be well approximated as being in a quasi-equilibrium
state, interest in stationary models of these stars arise in order to
understand the physical mechanism stabilizing the hypermassive star
without performing a complete numerical merger simulation.

Stationary models of (cold) relativistic stars have been extensively
explored in the literature. The first models, the famous
Tolman-Oppenheimer-Volkoff solutions, describing spherically symmetric
(therefore non-rotating) stars, date from the late
1930's~\cite{Tolman1939,Oppenheimer1939}. Hartle and Thorne proposed
the first axisymmetric rotating solutions from a perturbative approach
in the slow rotation approximation~\cite{HartleThorne}. Nowadays
several publicly available codes are able to obtain precise numerical
solutions up to the mass shedding limit~\cite{Nozawa1998,Ansorg2008},
at the Kepler frequency, see e.g. the textbook by
\citet{bookfriedmanstergioulas}. However, all these solutions only
treat cold $\beta$-equilibrated stars with a barotropic EoS. Goussard
et al.~\cite{Gouss1,Gouss2} have introduced the first models including
the effect of finite temperature, restricting their solution however
to the isentropic or isothermal case in $\beta$-equilibrium with
several fixed overall lepton fractions, where the EoS effectively
reduces to a barotropic one. Refs.~\cite{Martinon:2014uua,Camelio2016}
propose general solutions within a perturbative slow rotation
approach.  In this work, we follow Ref.~\cite{Villain2004} to
consistently compute stationary rapidly rotating hot stars based on
the publicly available numerical library
\textsc{lorene}~\cite{LORENE}.

The paper is organized as follows. In Sec.~\ref{sec:eos} we present
the new EoS model and in Sec.~\ref{sec:constraints} some of its
properties and in particular its compatibility with available
constraints. In Sec.~\ref{sec:structure} we present the formalism to
treat stationary rotating relativistic stars at nonzero
temperature. Sec.~\ref{sec:results} shows first applications of our
models, discussing maximum masses of hot stars and $I$-$Q$ relations,
i.e., universal relations among the moment of inertia and the
quadrupole moment. We conclude in
Sec.~\ref{sec:conclusions}. Throughout the paper we use natural units
with $c=\hbar= k_B=1$ where appropriate.

\section{Equation of state}
\label{sec:eos}
Although the transition to the quark-gluon plasma is very interesting,
as it could facilitate the supernova explosion~\cite{sagert_09},
explain some gamma-ray bursts~\cite{pili_16}, or -- within the
scenario of ``quark-novae'' -- some unusual supernova
lightcurves~\cite{Ouyed2002,Ouyed2013,Buballa_14}, we will concentrate
here on hyperonic degrees of freedom. Presently available general
purpose EoS models including all hyperons and covering the entire
range in baryon number density, $n_B$, temperature $T$ and hadronic
charge fraction, $Y_Q = n_Q/n_B = Y_e$~\footnote{$n_Q$ represents the
  total hadronic charge density.}  necessary for applications in CCSN
or binary mergers, are either not compatible with some constraints
from nuclear physics and/or a neutron star maximum mass of
$2 M_\odot$, see e.g.~\cite{ishizuka_08,Oertel:2012qd} or consider
only $\Lambda$-hyperons (e.g.~\cite{Banik:2014qja}). Our new EoS,
taking into account the entire baryon octet, is well compatible with
the main present constraints, see Sec.~\ref{sec:constraints} for
details.

\subsection{Statistical model for inhomogeneous matter}
At subsaturation densities and low temperatures, nucleonic matter is
unstable with respect to variations in the particle densities and
becomes inhomogeneous, i.e. nuclei or more generally nuclear clusters
are formed. The critical temperature is of the order $\sim 15$ MeV
just below saturation and decreases to about 1 MeV at lower densities.
Below a density of roughly $n_B \sim 10^{-4} \mathrm{fm}^{-3}$, the
cluster size is very small compared with its mean free path, such that
matter can be described as a non-interacting gas of nuclei, nucleons
and leptons in thermodynamic equilibrium. This approach is generally
called ``nuclear statistical equilibrium'' (NSE). In the last years
several models have been developed to go beyond a pure NSE and take
into account nucleon interactions and the interaction of clusters with
the surrounding medium at
higher densities (see e.g.~\cite{Raduta:2010ym,Gulminelli:2015csa,
  Hempel09,Typel2009, Sumiyoshi08a, Buyukcizmeci_14, Heckel09}). In
stellar matter particular attention has to be paid to the interplay
between the short-range nuclear interaction and the long-range Coulomb
interaction, which determines sizes and shapes of the nuclear clusters
and influences thus strongly the transition to homogeneous
matter~\cite{Gulminelli_03,Gulminelli_12}.

In the present EoS, clustered matter is described within the extended
NSE model of Hempel \& Schaffner-Bielich~\cite{Hempel09,hempel12}. Nuclei are
treated as classical Maxwell-Boltzmann particles. For the description
of nucleons, a relativistic mean field (RMF) approach is employed (see
Section \ref{sec:homogeneous} for details) with the same
parameterization as for the description of homogeneous matter. Several
thousands of nuclei are considered, including light ones other than
the $\alpha$-particle. If available, nuclear binding energies are
taken from experimental measurements \cite{audi_03}. In particular for
neutron rich nuclei, where no measurement exists, they are
complemented with values from theoretical nuclear structure
calculations \cite{moeller_95}. Several corrections are considered to
describe the modifications of cluster properties in medium: screening
of the Coulomb energies by the surrounding gas of electrons, excited
states, and excluded-volume effects.

\subsection{Homogeneous matter}
\label{sec:homogeneous}
Homogeneous matter is described within a phenomenological RMF. The
basic idea of this type of models is that the interaction between
baryons is mediated by meson fields inspired by the meson exchange
models of the nucleon-nucleon interaction. Within RMF models, these
are, however, not real mesons, but introduced on a phenomenological
basis with their quantum numbers in different interaction
channels. The coupling constants are adjusted to a chosen set of
nuclear observables. Earlier models introduce non-linear
self-couplings of the meson fields in order to reproduce correctly
nuclear matter saturation properties, whereas more recently
density-dependent couplings between baryons and the meson fields have
been widely used. The literature on those models is large and many
different parameterizations exist (see e.g.~\cite{Dutra2014}).

In the present paper, we will use models with density dependent
couplings. The Lagrangian density can be written in the following
form~\footnote{Note that we work here with a locally flat Minkowski
  metric $\eta^{\mu\nu}$. For the
  $\gamma$-matrices, we use the anticommutation relation
  $\{\gamma^\mu,\gamma^\nu\} = 2\, \eta ^{\mu\nu}$. }
%%%%%%%%%%%%%%%%%%%%%%%%%%%%%%%%%
\begin{eqnarray} {\mathcal L} &=& \sum_{j \in \mathcal{B}} - \bar
  \psi_j \left( \gamma_\mu \partial^\mu + m_j - g_{\sigma j} \sigma -
    g_{\sigma^* j} \sigma^* \right.\nonumber \\ &&\left. -
    i\,g_{\omega j} \gamma_\mu \omega^\mu - i \,g_{\phi j} \gamma_\mu
    \phi^\mu - i \, g_{\rho j} \gamma_\mu \vec{\rho}^\mu \cdot
    \vec{I}_j\right) \psi_j \nonumber \\ && - \frac{1}{2}
  (\partial_\mu \sigma \partial^\mu \sigma + m_\sigma^2 \sigma^2)
  \nonumber \\ && - \frac{1}{2} (\partial_\mu \sigma^* \partial^\mu
  \sigma^* + m_{\sigma^*}^2 {\sigma^*}^2) \nonumber \\ && -
  \frac{1}{4} W^\dagger_{\mu\nu} W^{\mu\nu} - \frac{1}{4}
  P^\dagger_{\mu\nu} P^{\mu\nu} - \frac{1}{4} \vec{R}^\dagger_{\mu\nu}
  \cdot \vec{R}^{\mu\nu} \nonumber \\ && - \frac{1}{2} m^2_\omega
  \omega_\mu \omega^\mu \nonumber \\ && - \frac{1}{2} m^2_\phi
  \phi_\mu \phi^\mu - \frac{1}{2} m^2_\rho \vec{\rho}_\mu \cdot
  \vec{\rho}^\mu ~,
\end{eqnarray}
where $\psi_j$ denotes the field of baryon $j$, and $W_{\mu\nu},
P_{\mu\nu}, \vec{R}_{\mu\nu}$ are the field tensors of the vector
mesons, $\omega$ (isoscalar), $\phi$ (isoscalar), and $\rho$
(isovector), of the form
\begin{eqnarray}
V^{\mu\nu} = \partial^\mu  V^\nu - \partial^\nu V^\mu~.
\end{eqnarray}
$\sigma, \sigma^*$ are scalar-isoscalar meson fields, coupling to all
baryons ($\sigma$) and to strange baryons ($\sigma^*$),
respectively. Some models introduce an additional scalar-isovector
coupling via a $\vec{\delta}$-meson, which we do not consider
here. The values of the baryon masses $m_j$ are chosen as follows:
$m_n = 939.565346, m_p = 938.272013, m_\Lambda = 1115.683, m_\Sigma =
1190, m_{\Xi^-} = 1321.68 ,m_{\Xi^0} = 1314.83$ MeV.

In mean field approximation, the meson fields are
replaced by their respective mean-field expectation values, which are given in
uniform matter as
\begin{eqnarray}
m_\sigma^2 \bar\sigma &=& \sum_{j \in B} g_{\sigma j} n_j^s
\\
m_{\sigma^*}^2 \bar\sigma^* &=& \sum_{j \in B} g_{\sigma^* j} n_j^s\\
m_\omega^2 \bar\omega &=& \sum_{j \in B} g_{\omega j} n_j\\
m_\phi^2 \bar\phi &=& \sum_{j \in B} g_{\phi j} n_j\\
m_\rho^2 \bar\rho &=& \sum_{j \in B} g_{\rho i} t_{3 j} n_j~,
\end{eqnarray}
where $\bar\rho=\langle\rho_3^0\rangle$,
$\bar\omega=\langle\omega^0\rangle$, $\bar \phi=\langle\phi^0\rangle$,
and $t_{3 j}$ represents the third component of isospin of baryon $j$
with the convention that $t_{3 p} = 1/2$. The scalar density of baryon
$j$ is given by
\begin{equation}
  n^s_j = \langle \bar \psi_j \psi_j \rangle = \frac{1}{\pi^2} \int
  k^2 \frac{M^*_j} {\sqrt{k^2 + M^{*2}_j}} \{f[\epsilon_j(k)]
  +\bar{f}[ \epsilon_j(k)] \} dk~,
\end{equation}
and the number density by
\begin{equation}
n_j = i\,\langle \bar \psi_j\gamma^0 \psi_j \rangle = \frac{1}{\pi^2} \int
k^2 (f[\epsilon_j(k)] - \bar{f}[\epsilon_j(k)])dk ~.
\end{equation}
$f$ and $\bar{f}$ represent here the occupation numbers of the
respective particle and antiparticle states with $\epsilon_j(k) =
\sqrt{k^2 + M^{*2}_j}$, and effective chemical potentials
$\mu^*_j$. They reduce to a step function at zero
temperature. The effective baryon mass $M^*_j$ depends on the scalar
mean fields as
\begin{equation}
M^*_j = M_j - g_{\sigma j} \bar\sigma - g_{\sigma^* j} \bar\sigma^*~,
\end{equation}
and
the effective chemical potentials
are related to the chemical potentials via
\begin{equation}
\mu_j^* = \mu_j - g_{\omega j} \bar\omega - g_{\rho j} \,t_{3 j}
\bar\rho - g_{\phi j} \bar \phi - \Sigma_0^R~.
\label{mui}
\end{equation}
The rearrangement term $\Sigma_0^R$ is present in models with
density-dependent couplings of meson $M$ to baryon $j$,
\begin{equation}
  g_{Mj}(n_B) = g_{Mj}(n_0) h_M(x)~,\quad x = n_B/n_0~,
\end{equation}
 to ensure thermodynamic consistency. It is given by
\begin{eqnarray}
  \Sigma_0^R &=& \sum_{j \in B} \left( \frac{\partial g_{\omega j}}{\partial n_j}
                 \bar\omega n_j + t_{3 j} \frac{\partial g_{\rho j}}{\partial n_j}
                 \bar\rho n_j +\frac{\partial g_{\phi j}}{\partial n_j}
                 \bar\phi n_j \right. \nonumber \\
             && \left. -\frac{\partial g_{\sigma j}}{\partial n_j}
                \bar\sigma n_j^s -\frac{\partial g_{\sigma^* j}}{\partial n_j}
                \bar\sigma^* n_j^s \right)~.
\end{eqnarray}
The density $n_0$ is a normalization constant, usually taken to be the
saturation density $n_0 = n_{\mathit{sat}}$ of symmetric nuclear
matter.

In the present paper we will consider the DD2
parameterization~\cite{Typel2009}, where the following form for the
density dependence of the isoscalar couplings is
assumed~\cite{Typel2009},
\begin{equation}
h_M(x) = a_M \frac{1 + b_M ( x + d_M)^2}{1 + c_M (x + d_M)^2}
\end{equation}
and
\begin{equation}
h_M(x) = a_M\,\exp[-b_M (x-1)] - c_M (x-d_M)~.
\end{equation}
for the isovector ones. The values of the parameters $a_M, b_M, c_M,$
and $d_M$ are listed in Ref.~\cite{Typel2009}.

Similar to many recent
works~\cite{Weissenborn11c,Banik:2014qja,Miyatsu:2013hea}, for the
hyperonic coupling constants, we will follow a symmetry inspired
procedure.  The individual isoscalar vector meson-baryon couplings are
expressed in terms of $g_{\omega N}$ and a few additional parameters,
$\alpha, \theta, z = g_1/g_8$, see e.g.~\cite{Schaffner96}, as follows
{\small
\begin{eqnarray}
\frac{g_{\omega\Lambda}}{g_{\omega N}} &=& \frac{ 1 - \frac{2 z}{\sqrt{3}}
  (1-\alpha) \tan \theta}{1- \frac{z}{\sqrt{3}}  (1 - 4 \alpha)
  \tan\theta} ~,\
\frac{g_{\phi\Lambda}}{g_{\omega N}} = -\frac{ \tan\theta + \frac{2 z}{\sqrt{3}}
  (1-\alpha)}{1- \frac{z}{\sqrt{3}}  (1 - 4 \alpha)
  \tan\theta} ~,\nonumber \\
\frac{g_{\omega\Xi}}{g_{\omega N}} &=& \frac{ 1 - \frac{z}{\sqrt{3}}
  (1+ 2 \alpha) \tan \theta}{1- \frac{z}{\sqrt{3}}  (1 - 4 \alpha)
  \tan\theta} ~,\
\frac{g_{\phi\Xi}}{g_{\omega N}} = -\frac{ \tan\theta + \frac{z}{\sqrt{3}}
  (1+ 2 \alpha)}{1- \frac{z}{\sqrt{3}}  (1 - 4 \alpha)
  \tan\theta} ~,\nonumber \\
\frac{g_{\omega\Sigma}}{g_{\omega N}} &=& \frac{ 1 + \frac{2 z}{\sqrt{3}}
  (1- \alpha) \tan \theta}{1- \frac{z}{\sqrt{3}}  (1 - 4 \alpha)
  \tan\theta} ~,\
\frac{g_{\phi\Sigma}}{g_{\omega N}} = \frac{ -\tan\theta + \frac{ 2 z}{\sqrt{3}}
  (1- \alpha)}{1- \frac{z}{\sqrt{3}}  (1 - 4 \alpha)
  \tan\theta} ~,\nonumber \\
\frac{g_{\phi N}}{g_{\omega N}} &=& -\frac{ \tan\theta + \frac{z}{\sqrt{3}}
  (1- 4 \alpha)}{1- \frac{z}{\sqrt{3}} (1 - 4 \alpha)
  \tan\theta} ~.
\label{eq:symmetry}
\end{eqnarray} }
Assuming an underlying $SU(6)$-symmetry, we will take $\tan\theta =
1/\sqrt{2}$, corresponding to ideal $\omega$-$\phi$-mixing, $\alpha =
1$, and $z = 1/\sqrt{6}$.
Extending the above procedure to the isovector sector would
lead to contradictions with the observed nuclear symmetry energy. $g_{\rho N}$
is therefore left as a free parameter and the remaining hyperonic isovector couplings are fixed by isospin symmetry.

The information from hypernuclear data on hyperonic single-particle
mean field potentials is then used to constrain the scalar coupling
constants.  The potential for particle $j$ in $k$-particle matter is
given by
\begin{equation}
U_j^{(k)}(n_k) = M^*_j - M_j + \mu_j - \mu^*_j~.
\end{equation}
We will assume here standard
values~\cite{Weissenborn11c,Banik:2014qja,Fortin2017} in symmetric
nuclear matter at saturation density, $n_{\mathit{sat}}$:
$U_\Lambda^{(N)}(n_{\mathit{sat}})= -30$ MeV,
$U_{\Xi}^{(N)}(n_{\mathit{sat}}) = -18$ MeV, and $U_\Sigma^{(N)}
(n_{\mathit{sat}}) = + 30$ MeV. The resulting values are in the range
obtained by calculating directly properties of single $\Lambda$
hypernuclei, see Refs.~\cite{vandalen_14,Fortin2017}.

Apart from a few light double-$\Lambda$-hyper\-nuclei, that constrain
only the low density behavior, almost no information is available on
the hyperon-hyperon ($YY$)-interaction and the corresponding
couplings, in particular $\sigma^*$ and $\phi$, are only very poorly
constrained. As mentioned above, we fix the $\phi$-couplings via the
relations in Eqs.~(\ref{eq:symmetry}) and neglect $\sigma^*$ for
simplicity in the main version of our EoS, named ``DD2Y''
hereafter. Without the coupling to $\sigma^*$, the $YY$-interaction is
very repulsive already at low densities. We obtain
$U_\Lambda^{(\Lambda)}(n_{\mathit{sat}}/5) = 7$~MeV,
$U_\Xi^{(\Xi)}(n_{\mathit{sat}}/5) = 47$~MeV, and
$U_\Sigma^{(\Sigma)}(n_{\mathit{sat}}/5) = 26$~MeV, whereas the data
on double-$\Lambda$-hypernuclei suggest a weakly attractive potential
at least for $\Lambda$-hyperons,
$U_\Lambda^{(\Lambda)}(n_{\mathit{sat}}/5) \approx
-1$--$-5$MeV~\cite{Vidana01,Khan15,Fortin2017}.  Although, as shown e.g. in
\cite{Oertel:2014qza,Oertel16}, the $\sigma^*$ has only a weak
influence on the EoS and (proto)-neutron star properties, we include a
second version (named ``DD2Y$\sigma^*$'' hereafter) of the EoS with a
$\sigma^*$-coupling adjusted to have
$U_\Lambda^{(\Lambda)}(n_{\mathit{sat}}/5) = -0.4$~MeV,
$U_\Xi^{(\Xi)}(n_{\mathit{sat}}/5) = -0.4$~MeV, and
$U_\Sigma^{(\Sigma)}(n_{\mathit{sat}}/5) =
-0.4$~MeV. Table~\ref{tab:couplings} summarizes the values of the
scalar meson hyperon couplings in both models obtained from the
above described procedure.
\begin{table}
\begin{tabular}{l|cccccc}
\hline
Model & $R_{\sigma \Lambda}$ & $R_{\sigma^* \Lambda}$ &$R_{\sigma
                                                        \Sigma}$
  &$R_{\sigma^* \Sigma}$ &   $R_{\sigma \Xi}$ &$R_{\sigma^* \Xi}$ \\
DD2Y & 0.62 & 0 & 0.48 & 0 & 0.32 & 0\\
DD2Y$\sigma^*$ & 0.62 & 0.46 & 0.48 & 0.84 & 0.32 & 1.11\\
\end{tabular}
\caption{\label{tab:couplings} Coupling constants of the scalar mesons
  to different hyperons within the two models presented here,
  normalized to the $\sigma N$- coupling from the DD2 parameter set,
  i.e. $R_{M j} = g_{M j}/ g_{\sigma N}$.}
\end{table}
Note that the couplings to $\Lambda$ in model DD2Y are the same as in
the BHB$\Lambda\phi$ EoS~\cite{Banik:2014qja}, where exactly the same
procedure has been followed.

\subsection{Combining different parts of the EoS}
The HS(DD2) EOS contains the transition from inhomogeneous or
clusterized matter to uniform nucleonic matter. This is done via the
excluded volume mechanism, which suppresses nuclei around and above
nuclear saturation density. On top of that, for some thermodynamic
conditions a Maxwell construction over a small range in density is
necessary, for details see Ref.~\cite{hempel12}.

Here the situation is slightly more complicated, since homogeneous
matter might contain hyperons. In the simplest case, hyperons appear
within homogeneous (nucleonic) matter and it is sufficient to minimize
the free energy of the homogeneous system to decide upon the particle
content of matter. Such a situation occurs at low temperatures and
high densities.

%%%%%%%%%%%%%%%%%%%%%%%%%%%%%%%%%%%%%%%%%%%%%
\begin{figure*}
\includegraphics[width=1.0\textwidth]{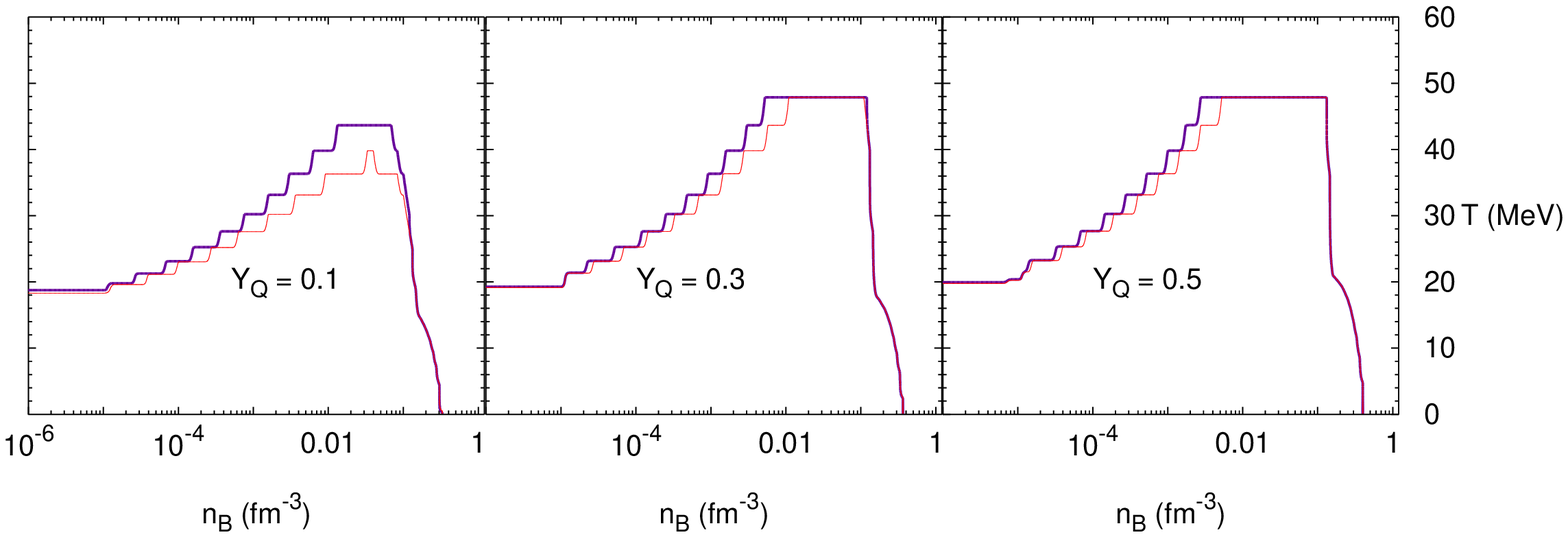}
\caption{(color online) The lines delimit the regions in temperature
  and baryon number density for which the overall hyperon fraction
  exceeds $10^{-4}$, which are situated above the lines. The dark thick
  purple line corresponds to the BHB$\Lambda\phi$ model and light thin red line
  to the DD2Y model. Different charge fractions are shown as
  indicated within the panels.
  \label{fig:contour}
}
\end{figure*}
%%%%%%%%%%%%%%%%%%%%%%%%%%%%%%%%%%%%%%%%%%%%%%%%%%%%%%%%%%%%%%%%%%%%%%%%%

In some parts of the $T$-$n_B$ diagram, however, a transition from
inhomogeneous matter directly to hyperonic homogeneous matter is
observed. This is the case at low densities and high temperatures,
i.e. the density regions up to the bumps in
Fig.~\ref{fig:contour}. There, light clusters compete with hyperonic
degrees of freedom with only very small differences in free energy
which are of the order of the numerical accuracy of the EoS
calculation. To technically construct the transition in this region,
we follow a similar prescription as in Ref.~\cite{Banik:2014qja} and
introduce a threshold value for the total hyperon fraction,
$Y_{\mathit{hyperons}} = \sum_{j \in B_Y} n_j/n_B$.  We let hyperonic
matter appear only if $Y_{\mathit{hyperons}} > 10^{-6}$. Note that the
hyperon fraction is not the same as the strangeness fraction, $Y_S$,
defined as the sum of all particle fractions multiplied by their
respective strangeness quantum numbers, $Y_S = \sum_{j \in B} S_j
n_j/n_B$.

Although the above described procedure allows to construct a smooth
transition between the different parts of the EoS, it is of course not
completely consistent. In principle, whenever hyperons compete with
light nuclear clusters, the free energy of the system should be
minimized allowing simultaneously for all different possibilities,
e.g. a coexistence of light clusters with hyperons. In view of the
tiny differences in free energy and the small fractions of particles
other than nucleons, electrons, and photons in the transition region,
a completely consistent treatment is left for future work.

\section{Equation of state properties}
\label{sec:constraints}
\subsection{Compatibility with constraints}
The interaction between nucleons can be constrained by data of finite
nuclei and nuclear matter properties. The latter are chosen in general
as the coefficients of a Taylor expansion of the energy per baryon of
isospin symmetric nuclear matter around saturation. Values with a
reasonable precision can be obtained for the saturation density
($n_{\mathit{sat}}$), binding energy ($E_B$), incompressibility ($K$),
symmetry energy ($E_{\rm sym}$) and its slope ($L$). In addition, much effort
has been recently devoted to theoretical ab-initio calculations of
pure neutron matter in order to constrain the equation of state. This
is particularly interesting for the EoS of compact stars, completing
the information about symmetric matter. The only robust
constraint on the interactions at super-saturation density arises from
the recent observation of two massive neutron stars, indicating the
maximum mass of a cold, non- or slowly-rotating (therefore
spherically symmetric) neutron star should be above $2 M_\odot$. A
summary and discussion of some of the most important available
constraints can be found e.g. in \cite{OertelRMP16}.

%%%%%%%%%%%%%%%%%%%%%%%%%%%%%%%%%%%%%%%%%%%%%
\begin{figure}
\includegraphics[width=1.0\columnwidth]{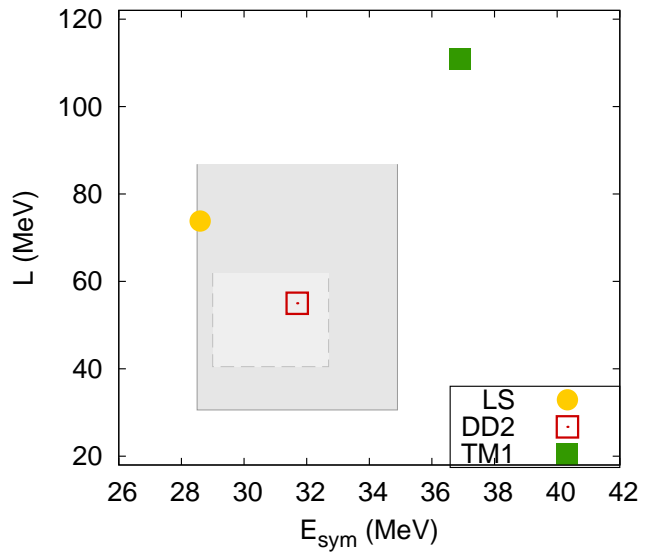}
\caption{(color online) Values of $E_{\rm sym}$ and $L$ in different nuclear
  interaction models. The two gray rectangles correspond to the range
  for $E_{\rm sym}$ and $L$ derived in Ref.~\cite{Lattimer:2012xj} (light Gray)
  and Ref.~\cite{OertelRMP16} (dark gray) from nuclear experiments and
  some neutron star observations.
  \label{fig:jl}}
\end{figure}
%%%%%%%%%%%%%%%%%%%%%%%%%%%%%%%%%%%%%%%%%%%%%%%%%%%%%%%%%%%%%%%%%%%%%%%%%
The present parameterization, DD2, has been chosen since it agrees
well with most of the established constraints. The values for
$n_{\mathit{sat}} = 0.149$~fm$^{-3}$, $E_B = 16.0$~MeV and $K =
243$~MeV are within standard ranges~\cite{OertelRMP16}. The
compatibility of $E_{\rm sym}$ and $L$ with ranges derived in
Ref.~\cite{Lattimer:2012xj} (light gray rectangle) and in
Ref.~\cite{OertelRMP16} (dark gray rectangle), respectively, are shown
in Fig.~\ref{fig:jl}. For comparison we show the values for two other
interactions, that of the Lattimer and Swesty EoS
(LS)~\cite{Lattimer:1991nc} and that for the TM1
parameterization~\cite{Sugahara_94}, too. These two interactions have
been employed in other recently developed general purpose EoS,
including non-nucleonic degrees of freedom,
e.g.~\cite{ishizuka_08,Shen:2011qu,Oertel:2012qd}.

%%%%%%%%%%%%%%%%%%%%%%%%%%%%%%%%%%%%%%%%%%%%%
\begin{figure*}
\includegraphics[width=1.0\textwidth]{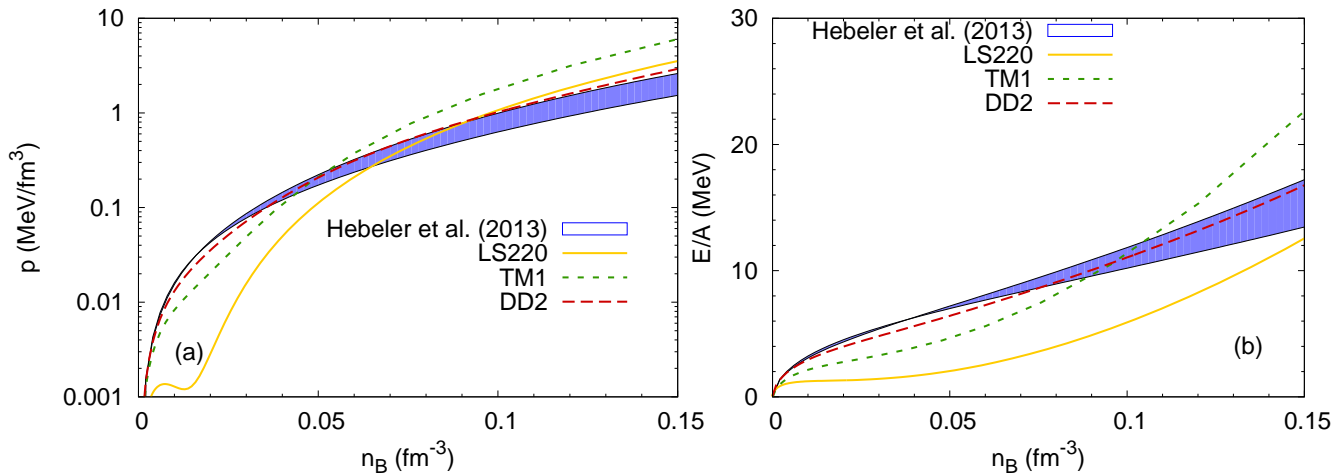}
\caption{(color online) Pressure (left panel) and energy per baryon
  (right panel) of pure neutron matter as functions of baryon number
  density within different nuclear interaction models compared with
  the ab-initio calculations of Ref.~\cite{Hebeler:2013nza}, indicated
  by the blue band.
\label{fig:nm}}
\end{figure*}
%%%%%%%%%%%%%%%%%%%%%%%%%%%%%%%%%%%%%%%%%%%%%%%%%%%%%%%%%%%%%%%%%%%%%%%%%
In Fig.~\ref{fig:nm} pressure and energy per baryon for pure neutron
matter are shown below saturation density. The blue band represents
the results from the ab initio calculations from
Ref.~\cite{Hebeler:2013nza} including an estimate of the corresponding
uncertainties. In contrast to LS and TM1, the interaction DD2 employed
here is in reasonable agreement with the ab initio calculations.

%%%%%%%%%%%%%%%%%%%%%%%%%%%%%%%%%%%%%%%%%%%%%
\begin{figure}
\includegraphics[width=1.0\columnwidth]{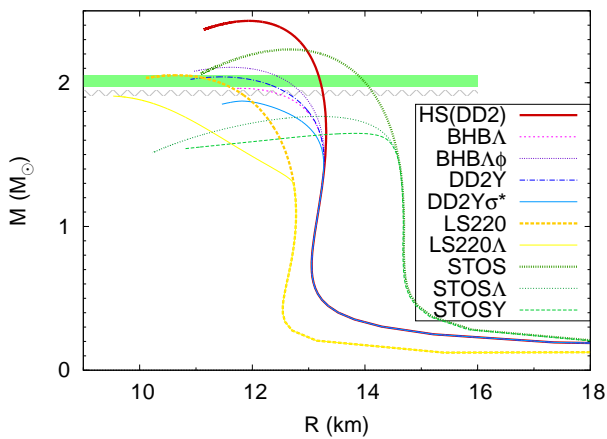}
\caption{(color online) Gravitational mass versus circumferential
  equatorial radius for cold spherically
  symmetric neutron stars within different EoS models. The two
  horizontal bars indicate the two recent precise NS mass
  determinations, PSR J1614-2230~\cite{Demorest2010,Fonseca2016}
  (hatched gray) and PSR J0348+0432~\cite{Antoniadis2013} (green).
\label{fig:mreos}}
\end{figure}
%%%%%%%%%%%%%%%%%%%%%%%%%%%%%%%%%%%%%%%%%%%%%%%%%%%%%%%%%%%%%%%%%%%%%%%%%
The mass-radius relation of cold\footnote{For convenience we have chosen a
    temperature of $T = 0.1$ MeV for producing this figure. In the
    following discussion of our results we always refer to this
    temperature upon speaking about ``cold'' stars.} 
 spherically symmetric neutron stars
within different general purpose EoS models is displayed in
Fig.~\ref{fig:mreos}. Purely nucleonic versions are shown with solid
lines, models including $\Lambda$-hyperons with dotted lines, and
those including the entire baryon octet with dashed-dotted lines. These
are the LS EoS~\cite{Lattimer:1991nc}, its extension with
$\Lambda$-hyperons (``LS220$\Lambda$'')~\cite{Peres_13}, the EoS by
Shen et al. (``STOS'') employing the TM1
interaction~\cite{Shen:1998by}, its extension with $\Lambda$-hyperons
(``STOS$\Lambda$'') \cite{Shen:2011qu} and all hyperons
(``STOSY'')~\cite{ishizuka_08}, as well as the two models including
$\Lambda$-hyperons within the same nuclear model as the present one
from Ref.~\cite{Banik:2014qja}, (``BHB$\Lambda$'') and
(``BHB$\Lambda\phi$''). It is evident from the figure that there are
only two EoSs including hyperons compatible with the $2
M_\odot$-constraint: BHB$\Lambda\phi$ containing only
$\Lambda$-hyperons and the present DD2Y. Both models are the
same, except for the particle content. The additional hyperonic
degrees of freedom in DD2Y slightly reduce the maximum mass with
respect to BHB$\Lambda\phi$, but it remains above $2M_\odot$. The
additional attractive $YY$-interaction in DD2Y$\sigma^*$ reduces the maximum
mass to 1.87 $M_\odot$, thus slightly below the observational limit. A summary
of cold neutron star properties for the different EoSs is given in
Table~\ref{tab:nsresultsT0}.

\subsection{Hyperon content and thermodynamic properties}
As already mentioned in Ref.~\cite{Oertel16}, the overall
hyperon content within the EoS remains similar between the models
containing only $\Lambda$-hyperons and the corresponding ones with the
full baryonic octet. For cold NSs, this can be seen from
Table~\ref{tab:nsresultsT0}. In Fig.~\ref{fig:contour}, the regions
where the overall hyperon fraction exceeds $10^{-4}$ are compared for
BHB$\Lambda$ and DD2Y. Although, as expected, hyperons are slightly
more abundant in the full model, the shape of the regions remains the
same and only small quantitative differences are observed. The bump in
the curves, i.e., the part of the lines above approximately 20~MeV, where
the abundance of hyperons is still below $10^{-4}$,
arises from the competition between light nuclear clusters
and hyperons in this particular temperature and density domain and
does not exist in the EoSs built on nuclear models without light
clusters, see Ref.~\cite{OertelRMP16}.

%%%%%%%%%%%%%%%%%%%%%%%%%%%%%%%%%%%%%%%%%%%%%%%%%%%%%%%%%%%%%%%%
\begin{table}
\begin{center}
\begin{tabular}{l|ccccc}
\hline
 Model& $M_g^{\mathit{max}}$  & $M_B^{\mathit{max}}$ & $R_{1.4}$ & $f_S$& $n_B^{(c)}$  \\
&$[M_\odot]$ & [$M_\odot$]& [km]& & [fm$^{-3}$] \\ \hline \hline
HS(DD2)         &2.43   & 2.90 & 13.27 & - &0.84 \\
BHB$\Lambda$  & 1.96  & 2.26&13.27  &0.05 & 0.95  \\
BHB$\Lambda\phi$  &2.11 &2.47 &13.27&0.05& 0.96\\
DD2Y & 2.04 &2.36 & 13.27 &0.04 &1.00\\
DD2Y$\sigma^*$ & 1.87 &2.15 & 13.27 &0.04 &0.98\\
\hline

\end{tabular}
\caption{Properties of cold spherically symmetric neutron stars in
  neutrinoless $\beta$-equilibrium: Maximum gravitational and baryonic
  masses, respectively, radius at a fiducial mass of
  $M_g = 1.4 M_\odot$, the total strangeness fraction, $f_S$,
  representing the integral of the strangeness fraction $Y_S/3$ over
  the whole star, defined as in Ref.~\cite{Weissenborn11c}, and the
  central baryon number density. The latter two quantities are given
  for the maximum mass configuration.  In addition to the EoSs
  presented here, for comparison the values for the purely nucleonic
  version HS(DD2)~\cite{Fischer_13} and the two versions including
  only $\Lambda$-hyperons from Ref.~\cite{Banik:2014qja} are listed.}
\label{tab:nsresultsT0}
\end{center}
\end{table}
%%%%%%%%%%%%%%%%%%%%%%%%%%%%%%%%%%%%%%%%%%%%%%%%%%%%%%%%%%%%%%%%%%%%

%%%%%%%%%%%%%%%%%%%%%%%%%%%%%%%%%%%%%%%%%%%%%
\begin{figure*}
\includegraphics[width=1.0\textwidth]{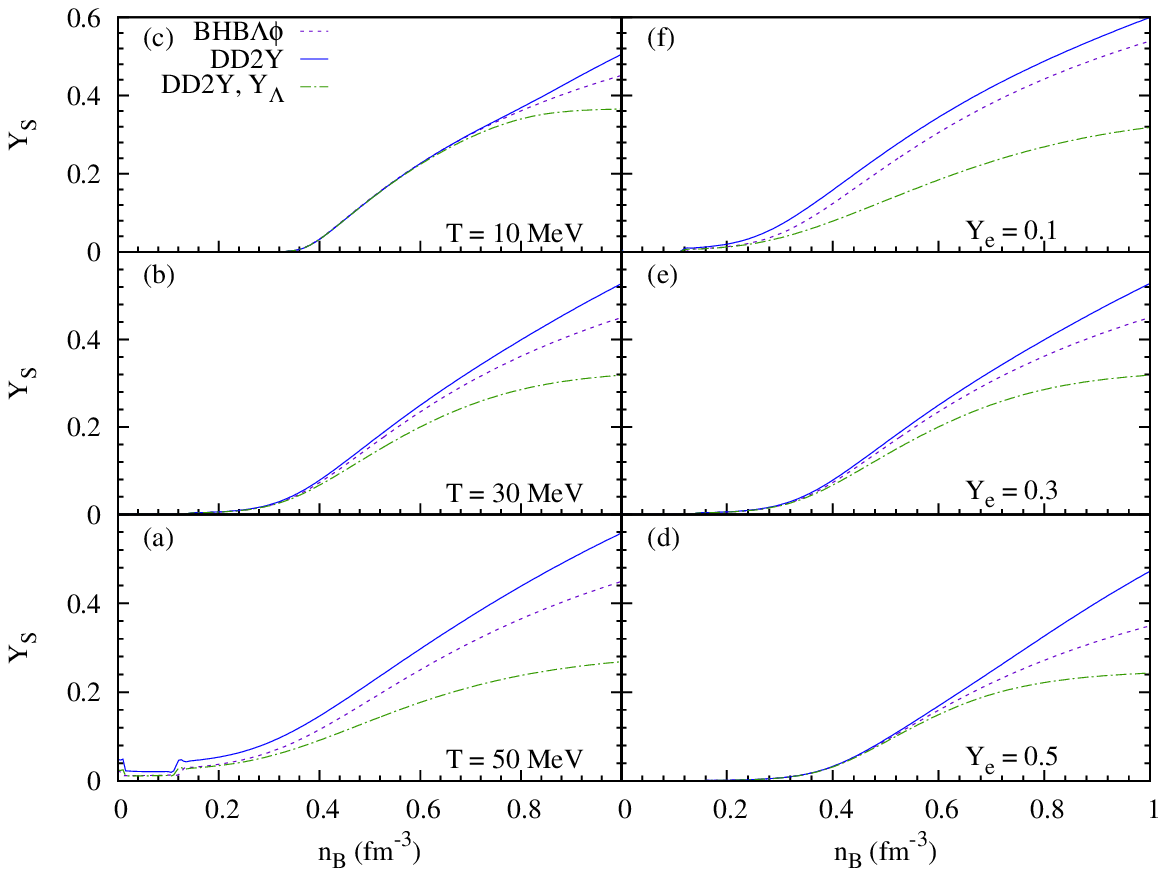}
\caption{(color online) Total strangeness fraction as function of
  baryon number density for different values of fixed temperature and
  electron fraction $Y_e$ within BHB$\Lambda\phi$ (dotted lines) and
  DD2Y (solid lines) EoS. In panels (a-c), $Y_e = 0.3$ and in panels
  (d-f), $T = 30$ MeV. For information, the $\Lambda$-fraction in
  model DD2Y is indicated, too (dash-dotted lines).
  \label{fig:hypfractions}
}
\end{figure*}
%%%%%%%%%%%%%%%%%%%%%%%%%%%%%%%%%%%%%%%%%%%%%%%%%%%%%%%%%%%%%%%%%%%%%%%%%
In Fig.~\ref{fig:hypfractions}, the overall strangeness fraction is
shown as function of baryon number density for different values of
fixed temperature and electron fraction for both models,
BHB$\Lambda\phi$ and DD2Y. As mentioned before, the hyperon onset
density remains similar in both models and the decrease in
$\Lambda$-fraction in DD2Y with respect to BHB$\Lambda\phi$ at high
densities is compensated by the presence of other hyperons such that
the overall strangeness fraction is larger in DD2Y. Note that here the
strangeness fraction $Y_S$ has been taken and not the hyperon
fraction. Naturally, the difference between both models increases with
increasing temperature. With increasing $Y_e$, as expected, in both
models the overall strangeness fraction decreases. The effect is,
however, less pronounced in DD2Y since the population of
neutral cascades and $\Sigma^+$ compensates
partially the suppression of other hyperonic degrees of freedom.
%%%%%%%%%%%%%%%%%%%%%%%%%%%%%%%%%%%%%%%%%%%%%
\begin{figure*}
\includegraphics[width=1.0\textwidth]{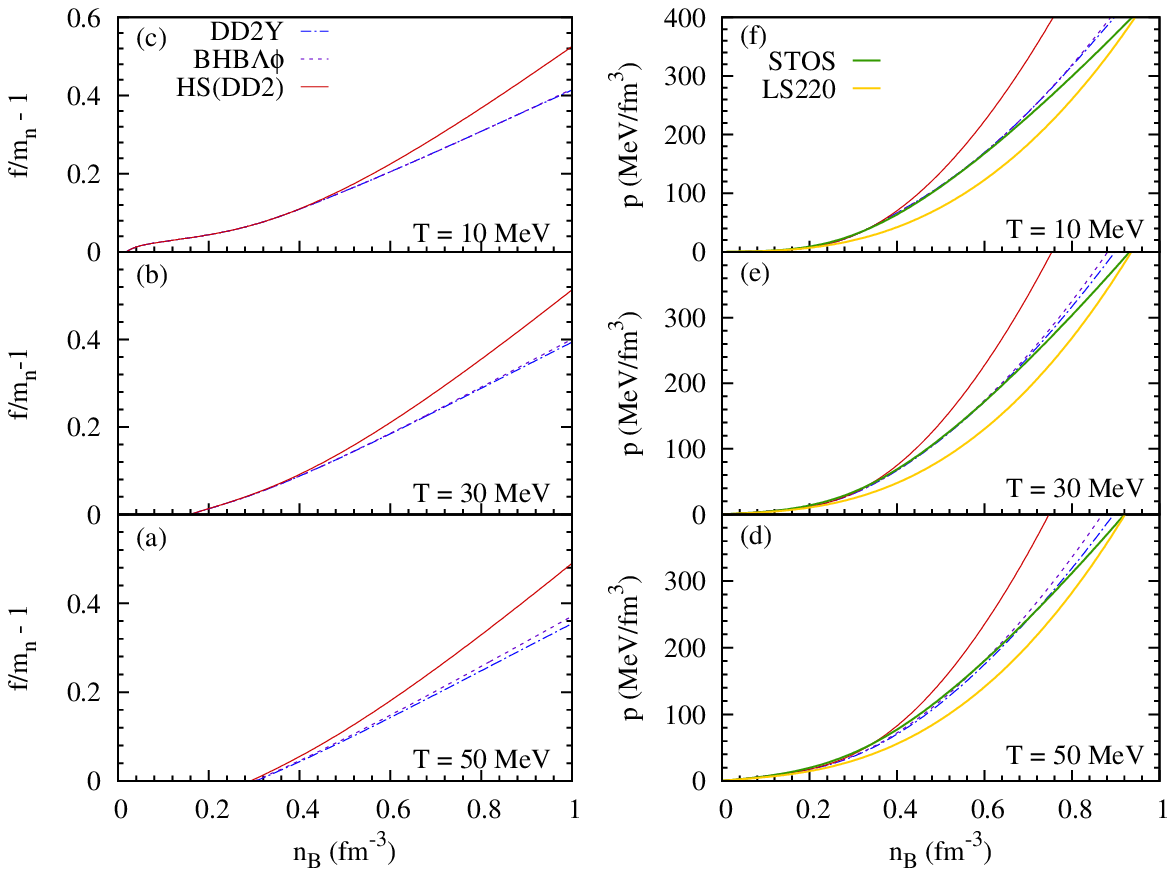}
\caption{(color online) Pressure (panels d-f) and normalised free
  energy per baryon (panels a-c) as function of baryon number density
  for different values of fixed temperature and $Y_e = 0.3$ within the
  three different EoS, HS(DD2), BHB$\Lambda\phi$ and DD2Y.  For
  information, the pressure in the classical models LS220 and STOS is
  displayed, too.
  \label{fig:hypthermo}
}
\end{figure*}
%%%%%%%%%%%%%%%%%%%%%%%%%%%%%%%%%%%%%%%%%%%%%%%%%%%%%%%%%%%%%%%%%%%%%%%%%

\begin{figure*}
\includegraphics[width=1.0\textwidth]{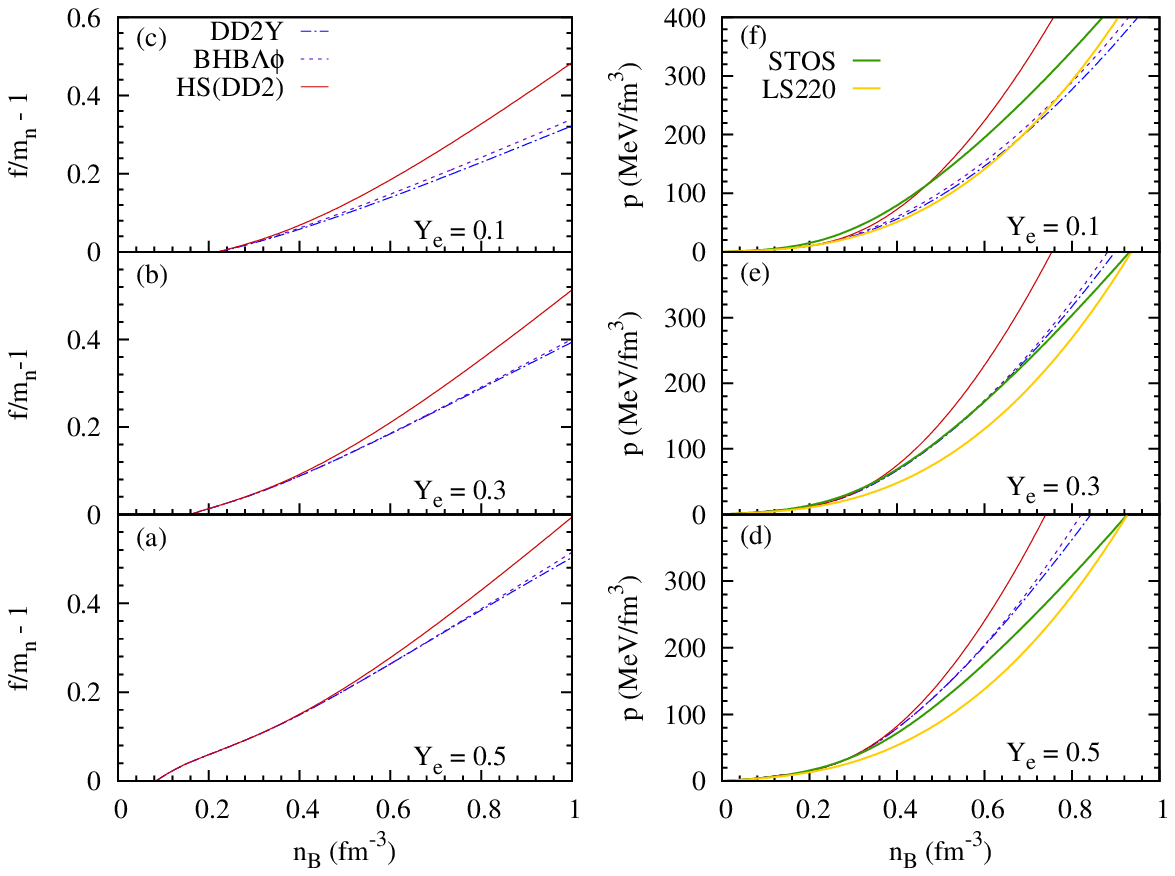}
\caption{(color online) Same as Fig.~\ref{fig:hypthermo}, but for $T = 30$ MeV and different fixed values of $Y_e$.
  \label{fig:hypthermoye}
}
\end{figure*}

Pressure and free energy per baryon are considerably reduced above
roughly 2-3 times nuclear saturation density in the models with
hyperons compared with the purely nucleonic HS(DD2) EoS, see
Figs.~\ref{fig:hypthermo} and \ref{fig:hypthermoye}. It is not
surprising that the reduction is most important for high temperatures
and low electron fractions. The presence of the full baryon octet in
DD2Y leads only to a small further reduction with repsect to the model
BHB$\Lambda\phi$, containing only $\Lambda$-hyperons. This is due to
the fact that the overall hyperon fraction is very similar in both
models, see the discussion above.
%%%%%%%%%%%%%%%%%%%%%%%%%%%%%%%%%%%%%%%%%%%%%
%%%%%%%%%%%%%%%%%%%%%%%%%%%%%%%%%%%%%%%%%%%%%%%%%%%%%%%%%%%%%%%%%%%%%%%%%

\section{Temperature dependent stellar structure}
\label{sec:structure}
In this section, we describe our strategy to solve for the star's
structure, following Ref.~\cite{Villain2004}. Equilibrium equations
will be solved together with Einstein equations, assuming stationarity
and axisymmetry. In addition, the matter content (represented by the
energy momentum tensor) should fulfill the circularity condition,
i.e. the absence of meridional convective currents. An EoS will close
the system of equations. In full generality the EoS depends on
temperature and on the different particle number densities or
thermodynamically equivalent variables. Conditions for electromagnetic
and strong equilibrium reduce the number of degrees of freedom in the
EoS to three, related to baryon number density $n_B$, electron number
density $n_e$ and temperature $T$. In neutron stars older than several
minutes, the temperature can be considered as vanishing and
neutrinoless weak $\beta$-equilibrium is achieved, such that the EoS
becomes effectively barotropic, i.e. depends only on baryon number
density or a thermodynamically equivalent variable. Neither in PNSs
nor in merger remnants these conditions are fulfilled and in particular a
nonzero temperature has to be considered.

Here, we will allow for an EoS with an explicit temperature
dependence. Under the current assumptions, in particular stationarity,
the most general solution for the star's structure becomes again
barotropic, and a relation $T(n_B)$ (or thermodynamically equivalent)
has to be provided~\cite{Gouss1,Gouss2,Villain2004}.  For simplicity
we will restrict the results within the present work either to
neutrinoless $\beta$-equilibrium or to a constant lepton fraction
$Y_L = (n_e + n_\nu)/n_B = n_L/n_B$ ($n_\nu$ and $n_L$ being,
respectively, neutrino and lepton number densities).  Following the
standard presentation of the formalism (see
e.g.~\cite{BGSM,Villain2004}), Latin letters $i,j, \dots$ are used for
spatial indices only, whereas Greek letters $ \alpha, \beta, \dots$
denote spacetime indices.

\subsection{Einstein equations}
General-relativistic models shall be described within the 3+1
formulation, where spacetime is foliated by a family of spacelike
hypersurfaces $\Sigma_t$, labeled by the time coordinate
$t$. Introducing coordinates ($x^i$) on each hypersurface, the line
element can be written as
\begin{equation}\label{e:3p1_metric}
ds^2=-N^2dt^2+\gamma_{ij}\left(dx^i+\beta^idt\right)\left(dx^j+\beta^jdt\right)~.
\end{equation}
$N$ represents the lapse function, $\beta^i$ the shift vector and
$\gamma_{ij}$ the 3-metric on each hypersurface $\Sigma_t$, thus
defining the spacetime metric $g_{\alpha\beta}$. More details can be found
e.g. in~\cite{3+1Eric}.

The assumptions of stationarity, axisymmetry and asymptotic flatness
imply the existence of two commuting Killing vector fields, given as
$\vec{\zeta} = \partial/\partial_t$ and $\vec{\chi} =
\partial/\partial_{\varphi}$ in an adapted coordinate system $(t, x^1,
x^2,\varphi)$. The two remaining coordinates are chosen to be spherical,
i.e. $x^1 = r, x^2 = \theta$. These adapted coordinates simplify the
expression of the metric: $\beta^r=\beta^\theta = 0$ and
$\gamma_{r\varphi} = \gamma_{\theta\varphi} = 0$. Finally, following~\cite{BGSM,
  Villain2004} we use a quasi-isotropic gauge, which additionally gives
$\gamma_{r\theta} = 0$, such that the line
element~(\ref{e:3p1_metric}) becomes
\begin{eqnarray}
  \label{e:QI_metric}
  ds^2&=&-N^2dt^2 + A^2\left(dr^2 + r^2 d\theta^2\right) \nonumber\\
  && + B^2r^2 \sin^2
  \theta \left( d\varphi^2 + \beta^\varphi dt\right)^2,
\end{eqnarray}
with the notations $A^2 = \gamma_{rr} = \gamma{\theta\theta}/r^2$ and
$B^2 = \gamma_{\varphi\varphi}/ (r^2 \sin^2\theta)$. All the metric
potentials $\left( N, \beta^\varphi, A, B \right)$ are functions of
the coordinates $(r,\theta)$ only. Einstein equations for these four
gravitational potentials, under our symmetry assumptions, reduce to a
set of four elliptic (Poisson-like) partial differential equations, in
which source terms contain both contributions from the energy-momentum
tensor (matter) and non-linear terms with non-compact support,
involving the gravitational field itself. Explicit expressions and
discussion of these equations can be found in~\cite{BGSM}.

\subsection{Equilibrium equations}

Matter is described as a perfect fluid with an energy-momentum tensor
of the form
\begin{equation}
 T^{\alpha\beta}=(\varepsilon+p)\, u^{\alpha}u^{\beta}+p\,g^{\alpha\beta}~.
\end{equation}
$\varepsilon$ denotes here the total energy density (including rest
mass), $p$ the pressure, and $u^\alpha$ is the fluid four-velocity;
the fluid angular velocity is then defined as $\Omega := u^\varphi/
u^t$. We also introduce the pseudo-log enthalpy
\begin{equation}
H = \ln\left({\frac{\varepsilon + p}{ m_B\,n_B}}\right)~,
\end{equation}
with $m_B$ a constant mass, where we chose the value $m_B= 939.565$ MeV.
Conservation of the energy-momentum
tensor\footnote{$\nabla_\alpha$ denotes here the covariant derivative
  associated to the 4-metric $g_{\alpha\beta}$}
$\nabla_\alpha T^{\alpha\beta} = 0$ yields the equation for the fluid
equilibrium~\cite{Gouss1,Gouss2}
\begin{equation}
\partial_i \left( H + \ln N - \ln \Gamma\right) =  \frac{T
  e^{-H}}{m_B} \partial_i s_B - u_\varphi u^t\partial_i\Omega~.
\label{finalDivT}
\end{equation}
$\Gamma = N u^t$ represents the Lorentz factor of the fluid with
respect to the Eulerian observer and $s_B$ the entropy
per baryon in units of the Boltzmann constant.

In this work, we will restrict ourselves only to the case where matter
is rigidly rotating ($\Omega=\textrm{const}$), which means that the
last term in Eq.~(\ref{finalDivT}) is zero. This equation is then
integrable in three cases. First, for a constant $s_B$, which is in
particular the case at zero temperature. The second case is the
isothermal one (constant $T^* = T N/\Gamma$) defined in
Ref.~\cite{Gouss1}. Finally, the most general solution in rigid
rotation is found introducing the heat function~\cite{Villain2004}
\begin{equation}
\hat{H}(n_B) = \int_0^{n_B} \frac{dp}{d n} \frac{1}{\varepsilon (n) + p(n)} dn~,
\label{eq:heatfunction}
\end{equation}
where a parameterization $T (n_B)$ has been assumed such that the EoS
effectively is again barotropic. Using $\hat{H}$, the equilibrium
condition reduces to:
\begin{equation}
  \label{e:equilibrium}
  \hat{H} + \ln N - \ln \Gamma = \textrm{const.},
\end{equation}
which is pretty similar to the zero-temperature case~\cite{BGSM}. It
is obvious that the heat function $\hat{H}$ reduces to the pseudo-log
enthalpy $H$ at zero temperature, up to a constant factor of
$\ln (m_B/\mu_B(n_B = 0))$, which can be absorbed in the r.h.s of
Eq.~(\ref{e:equilibrium}). For differentially rotating stars, allowing
the rotation law to depend on the entropy profile, in principle, the
condition of the EoS being barotropic could be relaxed. Such a scheme
is, however, beyond the purpose of the present paper.

We have implemented the above described scheme, with a temperature
dependent EoS within the numerical library \textsc{lorene}~\cite{LORENE},
see also refs.~\cite{BGSM,Villain2004}. The resolution of
elliptic-type partial differential equations (Einstein equations in
our case) is based on multi-domain spectral methods \cite{SpectralRev}
and is widely used for the computation of stationary rotating compact
objects. The equilibrium condition~(\ref{e:equilibrium}) is integrated
in a straightforward way and the heat function~(\ref{eq:heatfunction})
computed using the trapezoidal rule. Finally, we can use either an
analytic (polytropic type) EoS or a tabulated realistic one, which is
interpolated in a thermodynamically consistent way using the scheme by
Swesty~\cite{SwestyInt}.

Input parameters for a rigidly rotating neutron star model are: a
temperature vs. density profile, a prescription for the lepton
fraction (either $\beta$-equilibrium or constant $Y_L$), an EoS, a
central value for the heat function $\hat{H}(r = 0)$ and a value for
the rotation frequency $\Omega$. We can then compute the numerical
solution of all field equations described above and deduce global
quantities such as gravitational mass $M_g$ (from the asymptotic
behavior of the gravitational potential $N$), angular momentum $J$
(from the asymptotic behavior of the gravitational potential
$\beta^\varphi$), or circumferential equatorial radius (from the
integration of the line element~(\ref{e:QI_metric}) along the star's
equator). More details about these calculations can be found
in~\cite{BGSM}.

\section{Models of hot stars}
\label{sec:results}
Within this section we will discuss results for both non-rotating
(maximal masses, Sec.~\ref{ss:maxmass}) and rotating ($I$-$Q$ relations,
Sec.~\ref{ss:IQ}) stars with nonzero temperatures, employing different
microscopic EoSs exposed in the preceding sections. For the study of
their properties, $Y_Q$ will be fixed either by the condition of
$\beta$-equilibrium and assuming that neutrinos freely leave the
system, i.e. a vanishing electron lepton number chemical potential
\begin{equation}
\mu_{L} = 0~,
\end{equation}
or by fixing the electron lepton fraction $Y_{L} = 0.4$. This value
lies slightly above typical values obtained from simulations, see
e.g.~\cite{Pons:2000xf}. We have chosen it in order to maximize the
differences to the $\beta$-equilibrated case and thus show the maximal
effect we would expect from composition. Muons will not be considered,
although they might have a non-negligible influence on the EoS at the
very center of the PNS~\cite{Oertel:2012qd}. Neither of these
conditions might be very realistic, since the hydrodynamic evolution
should be coupled to neutrino transport, fixing the corresponding
evolution of $Y_e = Y_Q$ inside the star. A more complete study of PNS
evolution, combining our models with neutrino transport, is left for
future work.

\begin{figure*}
\includegraphics[width=0.99\columnwidth]{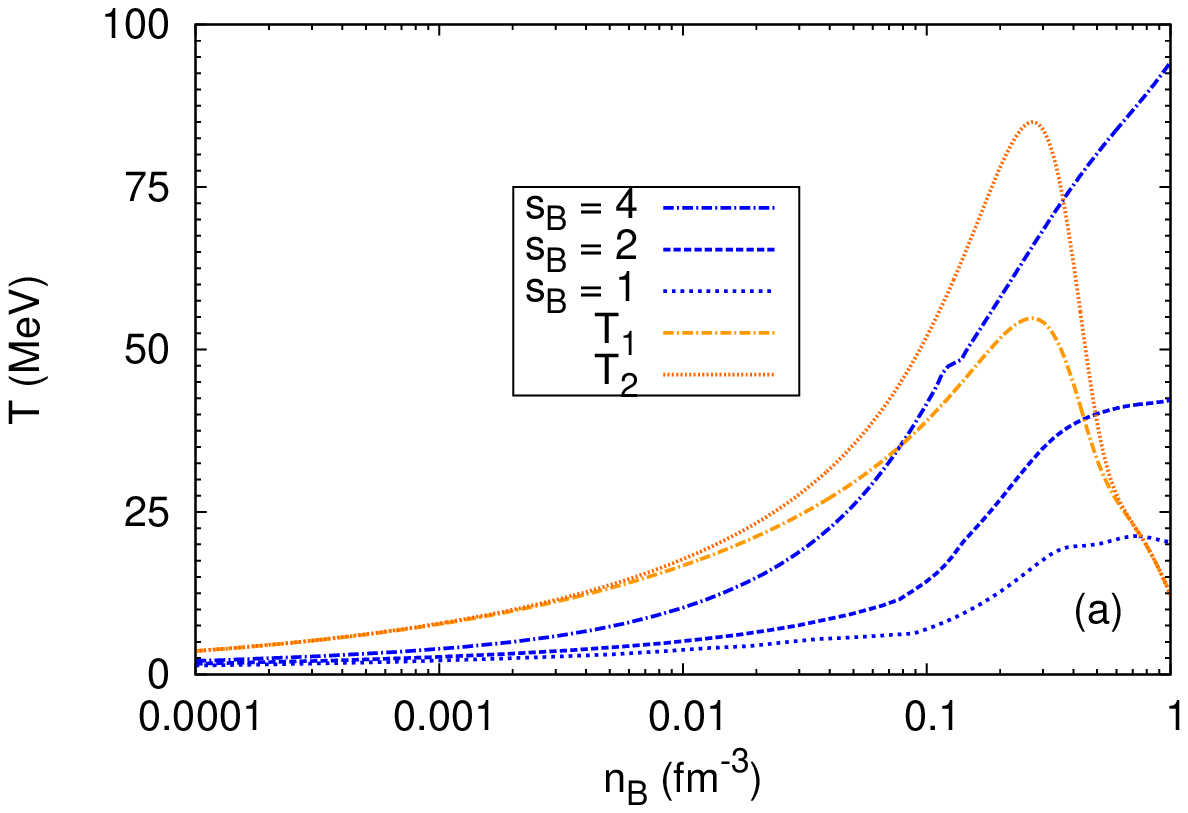}\hfill
\includegraphics[width=0.99\columnwidth]{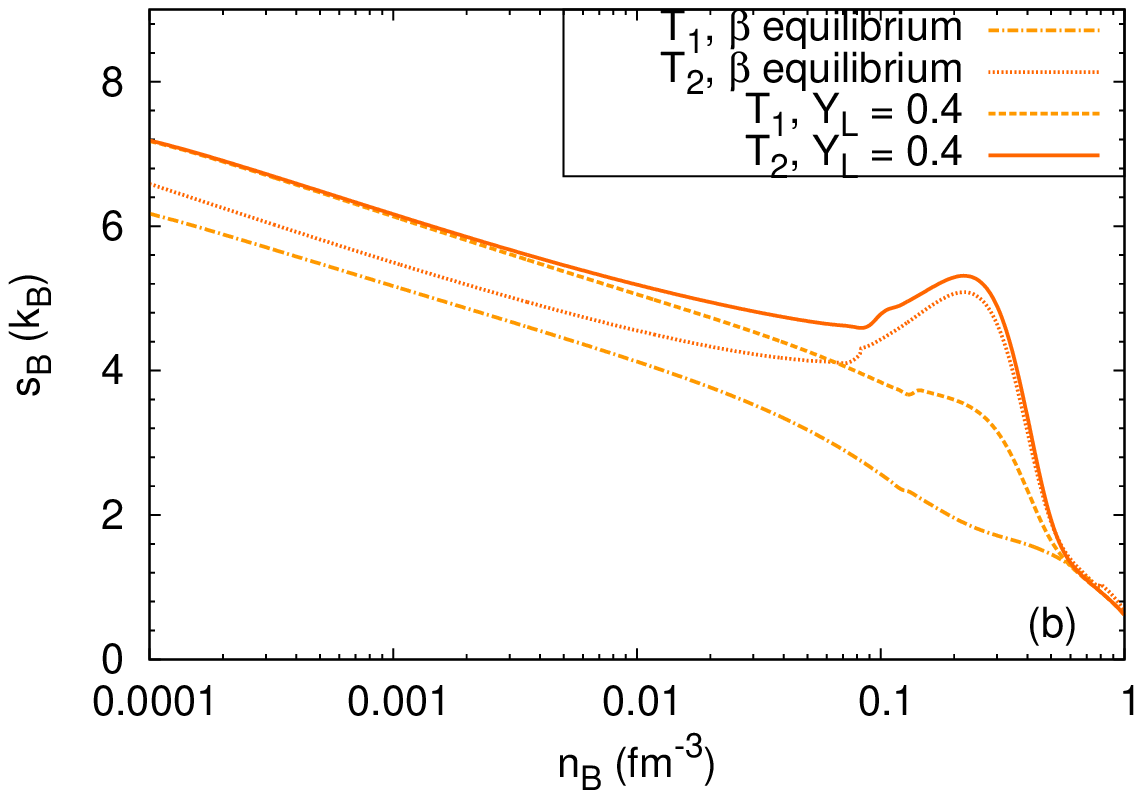}
\caption{Left: $T(n_B)$ relations used for the computation of hot star
  models. For comparison, the profiles obtained for different values
  of constant $s_B$ within the DD2Y EoS are shown, too. Right: $s_B(n_B)$ resulting from the two chosen $T(n_B)$ relations within the DD2Y EoS.
\label{fig:tn}}
\end{figure*}

Hence, we do certainly not pretend to give a completely realistic
picture of a PNS or a merger remnant. This simplified setup is
nevertheless sufficient for the purpose of the present study, namely
to demonstrate the usability of the newly developed EoS within a
numerical code, and to get some ideas about the influence of hyperons
on the properties of hot stars. Results with different temperature
profiles will be presented: either yielding constant values of entropy
per baryon, $s_B$, or profiles shown in Fig.~\ref{fig:tn} (left
panel), inspired by realistic calculations of PNS evolution. Profile
$T_1$ is within the range of values from the ``canonical'' simulations
of cooling PNSs~\cite{Pons1999}. The maximum temperatures of
profile $T_2$ is slightly above typical values from the aforementioned
simulations and corresponds to values reached for a PNS formed in the
collapse of a very massive progenitor star, close to the eventual
collapse to a black hole, see, e.g., Fig. 16~of
Ref.~\cite{hempel12}. Analytic expressions for both temperature
profiles are given in appendix~\ref{app:profiles}. The corresponding
entropy profiles are displayed in the right panel of
Fig.~\ref{fig:tn}.

%%%%%%%%%%%%%%%%%%%%%%%%%%%%%%%%%%%%%%%%%%%%%
%%%%%%%%%%%%%%%%%%%%%%%%%%%%%%%%%%%%%%%%%%%%%%%%%%%%%%%%%%%%%%%%%%%%%%%%%

\subsection{Maximal masses of non-rotating hot stars}\label{ss:maxmass}
The maximum baryonic mass a hot star can support is interesting both
for the merger remnant of a binary coalescence, and for the PNS after
the bounce occurs in a core collapse event, in order to determine the
conditions for the formation of a black hole. Different mechanisms
were evoked for stabilizing these objects against collapse to a black
hole. First, these objects are supposed to be rotating and, as
rotational effects on the maximum mass have been examined elsewhere,
see e.g. Ref.~\cite{Gouss1,Gouss2,Kastaun2015}, we do not discuss them
here. In addition, for the merger remnant and the PNS in the case of
collapse of fast spinning progenitor stars, the rotation profile is
strongly differential. Although it is not clear what are the time
scales driving toward rigid rotation, strong differential rotation can
help in supporting very massive configurations~\cite{Baumgarte1999,
  Morrison2004, Kastaun2015}.

Next, in PNSs, the lepton rich environment certainly contributes to
support a higher mass~\cite{Prakash1996, Pons1999} and it is not only
the cooling, but also the deleptonization via neutrino emission of the
star which causes a potential collapse to a black hole. In a merger
remnant, which is supposed to be close to $\beta$-equilibrium, this
mechanism cannot play the same role. Finally, canonical calculations
suggested that thermal pressure is unlikely to be able to to stabilize
the star~\cite{Prakash1996,Pons1999,Kaplan2013}, it might even
slightly reduce the maximum mass due to the population of additional
degrees of freedom at finite temperature. However, these studies were
restricted to rather low entropy values. For PNSs formed in
core-collapse of massive progenitors, which eventually are expected to
collapse to a black hole, it was found in
Refs.~\cite{hempel12,steiner13} that thermal effects can increase the
maximal gravitational mass by up to 0.6~$M_\odot$, where neutrinoless
$\beta$-equilibrium and a constant entropy per baryon of $s_B = 4$ was
considered.

When studying the maximum mass, previous works were considering cold
stars~\cite{salgado94,Morrison2004}, or a very restricted set of EoS,
containing only homogeneous matter~\cite{Prakash1996,Pons1999} or only
nucleonic matter~\cite{hempel12,steiner13,Kaplan2013,Kastaun2015}. Our
new EoS including hyperonic degrees of freedom allows to check the
influence of these new degrees of freedom on the mass, treating
consistently nuclear clustering at low densities and temperatures. A
recent study of PNSs with EoSs containing anti-kaons can be found in
Ref.~\cite{Batra2017}.

%%%%%%%%%%%%%%%%%%%%%%%%%%%%%%%%%%%%%%%%%%%%%%%%%%%%%%%%%%%%%%%%
\begin{table*}
\begin{center}
\begin{tabular}{|l||cc||ccc|ccc|ccc|ccc|}
\hline
& \multicolumn{2}{c||}{$S = 0 M_\odot$}
 & \multicolumn{3}{c|}{$S = 3 M_\odot$}  & \multicolumn{3}{c|}{$S = 5 M_\odot$}  & \multicolumn{3}{c|}{$S = 7 M_\odot$}
& \multicolumn{3}{c|}{$S = 9 M_\odot$}
\\ \hline
 & $M_g^{\mathit{max}}$ & $M_B^{\mathit{max}}$
 & $M_g^{\mathit{max}}$ & $M_B^{\mathit{max}}$& $T^{(c)}$ &
$M_g^{\mathit{max}} $& $M_B^{\mathit{max}}$& $T^{(c)}$ &
$M_g^{\mathit{max}}$ & $M_B^{\mathit{max}}$&$T^{(c)}$ &
$M_g^{\mathit{max}} $& $M_B^{\mathit{max}}$&$T^{(c)}$  \\
Model &
\multicolumn{2}{c||}{} &
\multicolumn{3}{c|}{}
& \multicolumn{3}{c|}{} & \multicolumn{3}{c|}{} & \multicolumn{3}{c|}{} \\
 & ($M_\odot$) & ($M_\odot$)
 & ($M_\odot$) & ($M_\odot$) & (MeV)&
($M_\odot$) & ($M_\odot$) &(MeV)&
($M_\odot$) & ($M_\odot$) &(MeV)&
($M_\odot$) & ($M_\odot$) &(MeV) \\
 \hline
HS(DD2)         &2.42 &2.90 & 2.43 &2.88&41  &2.43 &2.83& 68 & 2.45&2.77&90  &2.50 &2.73&108
\\
BHB$\Lambda\phi$  &2.11 &2.47 &2.11  &2.42&41 &2.13 &2.39& 67  &2.18 &2.37& 91 & 2.27&2.39&  107
\\
DD2Y &2.04 &2.35 &  &&  & &&  &2.02 &2.15&81 &2.11 &2.17& 96 \\
\hline
HS(DD2)         &- & - &2.37  &2.70 &32  &2.38 &2.67 & 53 &2.40 &2.64 & 71 &2.44 &2.61&89  \\
BHB$\Lambda\phi$  &- &- &2.17   &2.42 &27  &2.18  &2.39 &49  &2.21 &2.37 &65 &2.27 &2.36 &86 \\
DD2Y &- &- & 2.17 &2.43 &22  &2.16 &2.36 & 39 & 2.16 &2.30 &55  &2.20 &2.27 & 70  \\
\hline
\end{tabular}
\caption{Maximum gravitational and baryonic masses in units of solar
  mass for non-rotating stars and different values of constant total
  entropy. The central temperature of the maximum mass configuration
  is given, too. Since the entropy per baryon, $s_B$, is constant for
  each configuration, its value for the respective maximum mass
  configurations can be obtained simply by dividing $S$ by
  $M_B^{\mathit{max}}$, see Eq.~(\ref{eq:Stotal}).  The upper part
  assumes neutrinoless $\beta$-equilibrium and in the lower part
  $Y_L = 0.4$. For sake of an easier comparison the maximum masses in
  the cold $\beta$-equilibrated case are recalled in the first two
  columns. No values are given for DD2Y and the $\beta$-equilibrated
  case at $S=3 M_\odot$ and $S= 5 M_\odot$ since at high central
  densities the electron fraction lies below the limiting value of the
  table ($Y_e < 0.01$). The corresponding curves in
  Fig.~\ref{fig:mrhotthermal} do not show a maximum, we could thus not
  determine the maximum masses.}
\label{tab:nsresultsconsts}
\end{center}
\end{table*}
%%%%%%%%%%%%%%%%%%%%%%%%%%%%%%%%%%%%%%%%%%%%%%%%%%%%%%%%%%%%%%%%%%%%

%%%%%%%%%%%%%%%%%%%%%%%%%%%%%%%%%%%%%%%%%%%%%
\begin{figure*}
\includegraphics[width=0.99\textwidth]{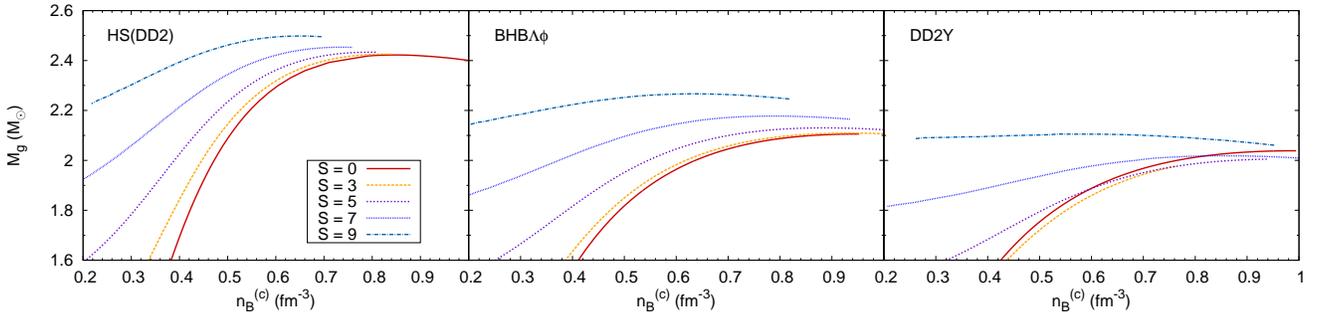}
\caption{Gravitational mass versus central baryon number density for
  non-rotating stars for three different EoSs, without hyperons (left),
  with $\Lambda$-hyperons (middle) and the complete baryon octet
  (right). Different values of constant total entropy $S$ have been
  used, indicated in units of solar masses. $\beta$-equilibrium has
  been assumed. For comparison, the cold result is shown,
  too. For DD2Y, at $S = 3 M_\odot$ and $S = 5 M_\odot$, the curves
  end at some central density above which no longer any
  $\beta$-equilibrated solution is found. The reason is that the
  electron fraction becomes lower than the limiting value of the EoS
  table ($Y_e = 0.01$). No maximum could be determined in this case.
\label{fig:mrhotthermal}}
\end{figure*}
%%%%%%%%%%%%%%%%%%%%%%%%%%%%%%%%%%%%%%%%%%%%%%%%%%%%%%%%%%%%%%%%%%%%%%%%%
In order to discuss maximum masses, we have to consider the stability
of the computed stellar configurations. At zero temperature for
non-rotating stars, stable configurations verify simply $d M_g/d
n_B^{(c)} \ge 0$. This criterion is a special case of the result by
Friedman et al.~\cite{Friedman88}, who have established a turning point
criterion for determining whether a rotating star becomes secularly
unstable with respect to axisymmetric perturbations. Here, we have
to use an extended version for hot (non-rotating) stars.

%%%%%%%%%%%%%%%%%%%%%%%%%%%%%%%%%%%%%%%%%%%%%
\begin{figure*}
\includegraphics[width=0.99\textwidth]{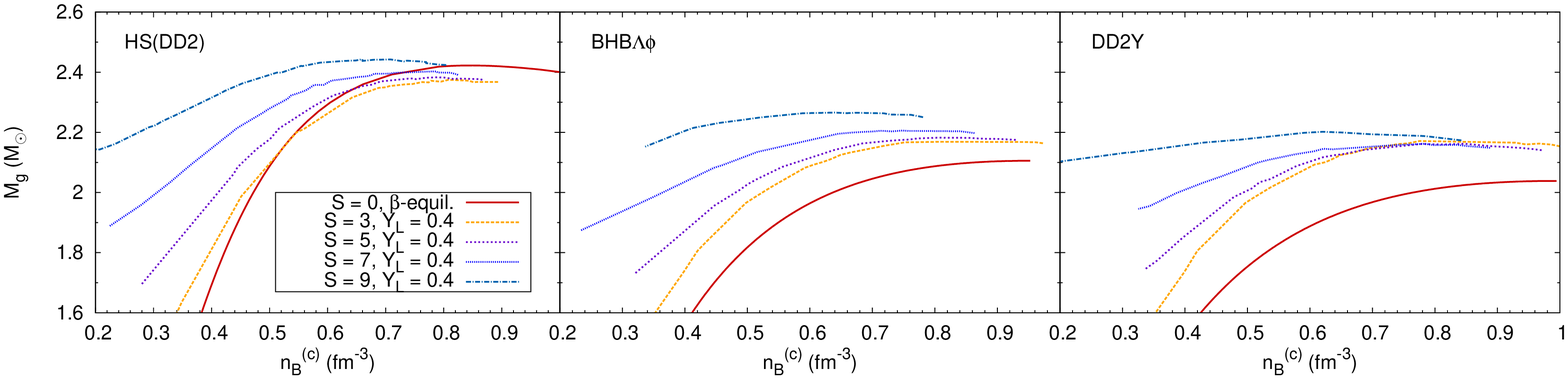}
\caption{Same as Fig.~\ref{fig:mrhotthermal}, comparing stars with
  $Y_L$ = 0.4 at different constant total entropy values with the
  $\beta$-equilibrated result for cold stars.
  \label{fig:mrhotcompo}}
\end{figure*}
%%%%%%%%%%%%%%%%%%%%%%%%%%%%%%%%%%%%%%%%%%%%%%%%%%%%%%%%%%%%%%%%%%%%%%%%%

For the following considerations, we will consider a more general case
for hot rotating stars and we will denote by $J$ the total angular
momentum of the star and $S$ the total entropy. $J$ is defined
as~\cite{BGSM}
\begin{equation}
J = \int A^2 B^2 (E + p) U r^3 \sin^2 \theta dr d\theta d\phi ~.
\end{equation}
$E$ denotes the energy density as measured by a locally non-rotating
observer, $E = \Gamma^2(\varepsilon + p) - p$, and $U$ the fluid
velocity as measured by the same observer. The latter is related to
the factor $\Gamma$ as $\Gamma = (1 - U^2)^{-1/2}$. $S$ can be
expressed in a similar way from
\begin{equation}
S = \int A^2 B \Gamma n_B s_B m_B r^2 \sin \theta dr d\theta d\phi ~.
\label{eq:Stotal}
\end{equation}
As shown in Ref.~\cite{Gouss1}, based on the work by
Sorkin~\cite{Sorkin82}, a meaningful criterion for a configuration
being secularly stable can be obtained for rigidly rotating stars with
a constant $s_B$ (or $T^*$) throughout the star. In the former case,
i.e. for constant $s_B$, the total entropy is simply given by
$S = s_B \, M_B$.  Following Ref. ~\cite{Gouss1}, a star becomes
unstable at the extremal points
\begin{equation}
\left ( \frac{\partial J}{\partial n_B^{(c)}} \right)_{M_B,S} = 0\; ,\;
\left ( \frac{\partial M_B}{\partial n_B^{(c)}} \right)_{J,S} = 0\; ,\;
\left ( \frac{\partial S}{\partial n_B^{(c)}} \right)_{M_B,J} = 0~.
\label{eq:criterion}
\end{equation}
Obviously, upon varying the central baryon number density (or
equivalently the central heat function) at constant angular momentum,
the rotation frequency changes, see e.g. the
textbook~\cite{bookfriedmanstergioulas}, i.e. sequences at constant
rotation frequency do not allow to distinguish stable from unstable
solutions. Equivalently, for sequences at constant total entropy, the
entropy per baryon $s_B$ is not constant, and sequences at constant
$s_B$ do not allow to identify stable and unstable configurations.

%%%%%%%%%%%%%%%%%%%%%%%%%%%%%%%%%%%%%%%%%%%%%
\begin{figure*}
\includegraphics[width=0.99\textwidth]{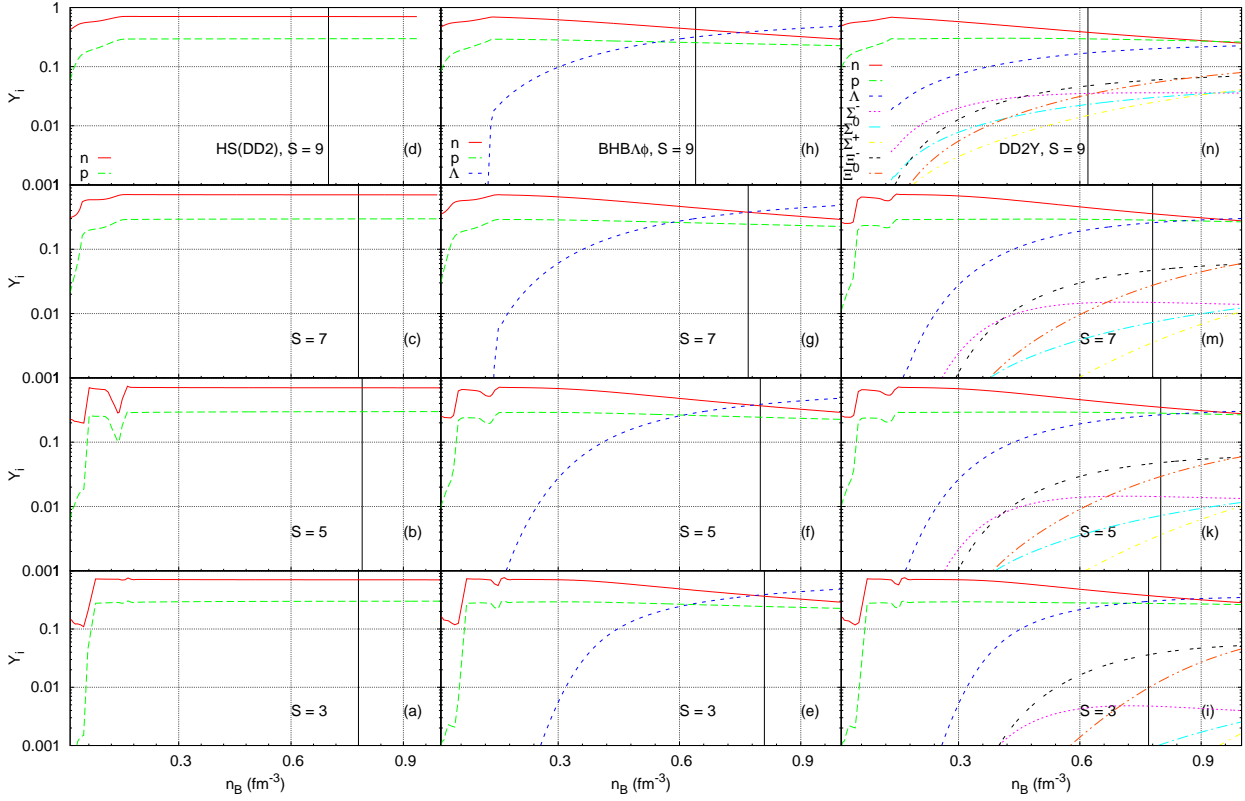}
\caption{
Particle fractions in hot neutron star matter for the
  three different EoS discussed here. The lepton fraction has been fixed to $Y_L = 0.4$ and the entropy per baryon corresponds to the value of the respective maximum mass configuration with the total entropy indicated in each panel. The vertical lines
  show the central density of the respective maximum mass configurations.
\label{fig:fractionsPNS}}
\end{figure*}
%%%%%%%%%%%%%%%%%%%%%%%%%%%%%%%%%%%%%%%%%%%%%%%%%%%%%%%%%%%%%%%%%%%%%%%%%
We are mainly interested here in thermal effects on the star's mass,
corresponding to the maximum mass a cooling star can support, i.e. the
second criterion of Eq.~(\ref{eq:criterion}) is the most interesting
one. It determines the maximum mass at different given values of
constant total entropy and angular momentum. In the following, we
restrict the discussion to non-rotating stars, for which $J=0$ and is
therefore constant. Since the criterion for distinguishing secularly
stable from unstable configurations is meaningful only for constant
$s_B$ (or $T^*$), we will restrict our investigations of maximum
masses to models with constant $s_B$, too.

The different values of the maximum mass are summarized in
Table~\ref{tab:nsresultsconsts}. In the upper part, neutrinoless
$\beta$-equilibrium is assumed, in the lower part a constant
$Y_L = 0.4$.

\subsubsection{Thermal effects}

In Fig.~\ref{fig:mrhotthermal} we thus display the gravitational
mass versus central baryon number density obtained for non-rotating
stars and different values of total entropy $S$.
Results for three
different EoSs are shown: the purely nucleonic one, HS(DD2), the one
containing $\Lambda$-hyperons, BHB$\Lambda\phi$, and the new EoS
considering the entire baryon octet, DD2Y. We do not show results for
DD2Y$\sigma^*$ here since it does not respect the cold neutron star
maximum mass constraint. $\beta$-equilibrium is assumed for all
calculations.

It is obvious that for $S = 3 M_\odot$, corresponding to
configurations with $s_B$ roughly between 1 and 2, thermal effects on
the maximum mass are small, and almost no difference can be observed
with respect to the result for cold stars. A slight reduction of the
gravitational mass for BHB$\Lambda\phi$ and, in particular, DD2Y, is
due to the population of additional degrees of freedom at finite
temperature for these two EoSs allowing for non-nucleonic
particles. These findings confirm previous investigations, see
e.g. Refs.~\cite{Prakash1996,Pons1999,Kaplan2013}. At low central
densities, thermal effects are more important. This can be understood
since for total entropy constant, with decreasing gravitational mass,
the entropy per baryon $s_B$ of the configurations increases, reaching
almost $s_B = 3$ at the lower end of the curves, modifying
considerably the EoS.

In contrast, at $S = 9$, thermal effects on the gravitational masses
are clearly non-negligible for all three EoS. The maximum mass is
increased by 4\% for HS(DD2), 8\% for BHB$\Lambda\phi$ and 7\% for
DD2Y, respectively. These values are of the same order as those
expected for rigid rotation~\cite{BGSM}. The temperatures and
entropies of these configurations are reached typically for PNSs in
the post-bounce phase of core-collapse events with massive
progenitors. The importance of thermal effects can be seen also from
the shift in central density of the maximum mass configurations
compared with the cold result. The central density is
reduced with increasing value of $S$ since the hot star becomes less
compact due to thermal excitations.

It should be pointed out that for a given entropy per baryon the
temperature is significantly lower within an EoS including hyperons
than in a purely nuclear one, see e.g.~\cite{Oertel16}. This is a
trivial thermodynamic effect: the appearance of hyperons implies that
the energy is shared among an increased number of degrees of freedom,
with consequently reduced thermal excitations for each of
them. Therefore, although the value of $s_B$ for the maximum mass
configurations is higher for the EoS with hyperons than for the purely
nuclear one, the central temperature with DD2Y is only 96 MeV, whereas
it is 108 MeV for HS(DD2).

\subsubsection{Composition}
It is known that a lepton rich environment disfavors hyperonic degrees
of freedom and that generally with increasing hadronic charge
fraction, $Y_Q$, the EoS becomes stiffer due to the reduced number
of degrees of freedom
present~\cite{Prakash1996,Pons1999,colucci_13,OertelRMP16,Oertel16}. Therefore
neutrino trapping has been evoked for a long time already as one of
the main mechanisms to stabilize a PNS with hyperons (or pions/kaons)
against collapse to a black hole. From Fig.~\ref{fig:mrhotcompo} it is
evident that our results confirm previous findings. We display the
gravitational mass of non-rotating stars as function of central baryon
number density for the three previously considered EoSs. A fixed
lepton fraction of $Y_L = 0.4$ and high temperatures ($S = 9$) or low
temperatures ($S = 3$) is compared with the respective
$\beta$-equilibrated results for cold stars.

As expected, for DD2Y, the lepton rich environment with $Y_L = 0.4$
clearly contributes to increasing considerably the gravitational mass
supported by the star. To less extent, this is true for
BHB$\Lambda\phi$, too. The difference between DD2Y and
BHB$\Lambda\phi$ becomes small since $\Lambda$-hyperons, being charge
neutral, are less affected by the higher electrons fraction than
charged hyperons, essentially $\Sigma^-$. In contrast, for the purely
nucleonic EoS HS(DD2) almost no difference between the lepton rich and
the $\beta$-equilibrated case is observed at $S = 3$ and only a
moderate increase for $S = 9$.  The combination of thermal and
composition effects leads to a maximum mass of $M_g = 2.2 M_\odot$
for DD2Y with $S = 9$ and $Y_L = 0.4$, $0.16 M_\odot (\approx 8 \%)$
above the cold $\beta$-equilibrated maximum mass.

It should be noted, however, that with increasing temperature the
hyperon suppression in a lepton rich environment becomes less
pronounced. Therefore, for DD2Y --and to less extent for
BHB$\Lambda\phi$, too-- the increase in maximum gravitational mass
with increasing total entropy is moderate at $Y_L = 0.4$. This can be
seen from Fig.~\ref{fig:fractionsPNS}, too, where the different
particle fractions for the maximum mass configurations are shown. For
$S = 9$, corresponding to $s_B = 3.96$ with DD2Y, all different
hyperonic species have non-negligible fractions at the center of the
star due to the high temperatures reached.
\subsection{$I$-$Q$-relation}\label{ss:IQ}
%%%%%%%%%%%%%%%%%%%%%%%%%%%%%%%%%%%%%%%%%%%%%
\begin{figure*}
\begin{center}
\includegraphics[width=0.7\textwidth]{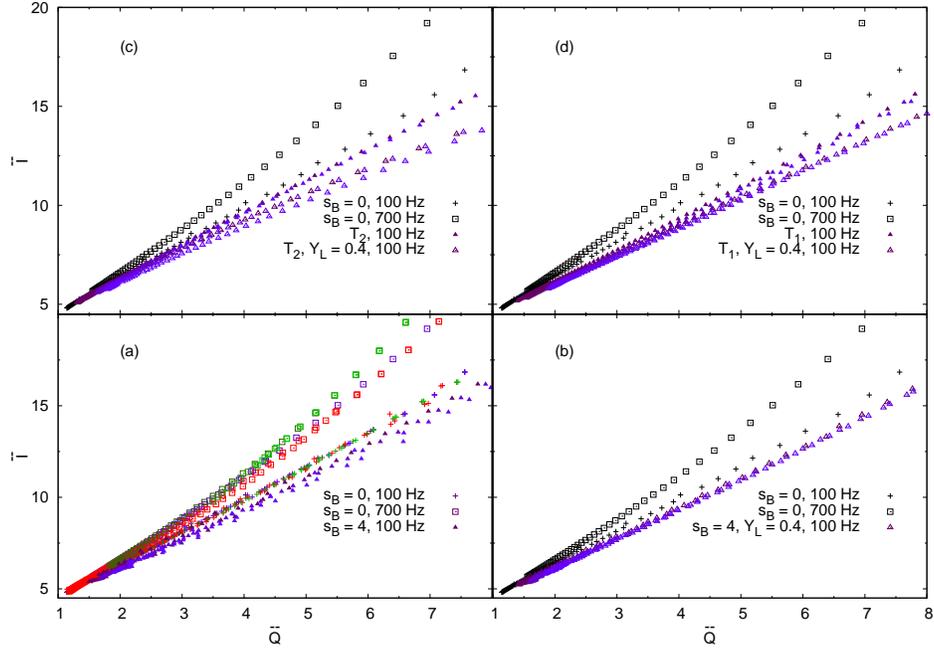}
\caption{(color online) Normalized moment of inertia versus quadrupole
  moment for different EoS. The color coding corresponds to different
  EoS models whereas different symbols indicate rotation frequencies
  and entropy per baryon of the stars. In panel (a) different
  nucleonic and hyperonic EoS are shown for cold stars, in
  $\beta$-equilibrium, for the slow as well as fast rotating case,
  respectively. The LS220 EoS and the counterpart with
  $\Lambda$-hyperons are thereby shown by red symbols, STOS EoS with
  and without hyperons by green symbols and the three EoSs based on
  DD2, HS(DD2), BHB$\Lambda\phi$ and DD2Y, by violet symbols. For the
  latter three EoS in addition the results for slow rotating
  configurations with $s_B = 4$ are displayed. In the three other
  panels the cold reference case with the HS(DD2) EoS is
  displayed by black symbols and different situations are considered
  with the three EoSs, HS(DD2), BHB$\Lambda\phi$, and DD2Y, indicated
  by violet symbols: constant $Y_L = 0.4$ (b), profile $T_2$ (c),
  profile $T_1$ (d).
\label{fig:iq}}
\end{center}
\end{figure*}
%%%%%%%%%%%%%%%%%%%%%%%%%%%%%%%%%%%%%%%%%%%%%%%%%%%%%%%%%%%%%%%%%%%%%%%%%

%%%%%%%%%%%%%%%%%%%%%%%%%%%%%%%%%%%%%%%%%%%%%
\begin{figure}
\begin{center}
\includegraphics[width=0.99\columnwidth]{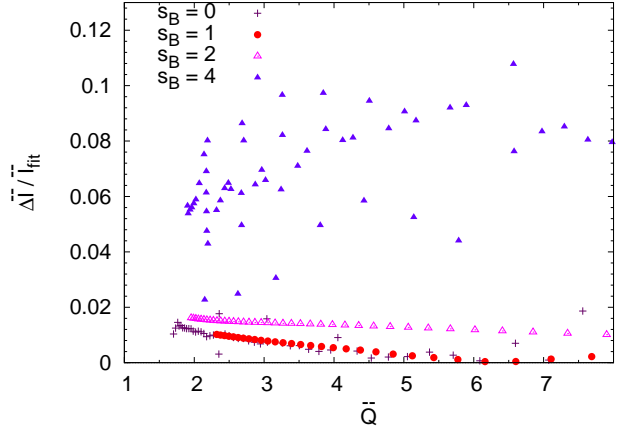}
\caption{(color online) Relative difference between
  Eq.~(\ref{eq:iqfit}) -- the fitted results for cold slowly rotating
  stars-- and the results at different constant $s_B$ values for the
  DD2Y EoS, assuming neutrinoless $\beta$-equilibrium.
\label{fig:iqfit}}
\end{center}
\end{figure}
%%%%%%%%%%%%%%%%%%%%%%%%%%%%%%%%%%%%%%%%%%%%%%%%%%%%%%%%%%%%%%%%%%%%%%%%%

It has been shown~\cite{Yagi:2013bca,Yagi:2013awa} that there exist
relations between the moment of inertia ($I$), the tidal deformability
($\lambda$) and the quadrupole moment ($Q$) of neutron stars which are
approximately independent of the internal composition and the EoS. Originally
proposed for slowly rotating cold neutron stars, they remain EoS
independent for fast rotation, too, and universal fits with a
functional form
\begin{equation}
\ln y = a + b\, \ln x + c \, (\ln x)^2 + d\, (\ln x)^3 + e\, (\ln x)^4
\label{eq:iqfit}
\end{equation}
can be established~\cite{Yagi:2013bca,Yagi:2013awa}. The coefficients
$a,b,c,d,e$ are frequency dependent~\cite{Doneva:2013rha} but do not
depend on the EoS. $x$ and $y$ represent any couple of the normalized
quantities $\bar I, \bar Q,\bar \lambda$
\begin{equation}
\bar I = \frac{I}{M^3_g} \quad ,\quad \bar Q = \frac{Q}{M^3_g
  (J/M^2_g)^2}\quad ,\quad \bar \lambda = \frac{\lambda}{M^5_g}~,
\end{equation}
with $M_g$ being the star's gravitational mass and $J$ its angular
momentum. We will employ here the numerical values for the fit
coefficients obtained from a fit to the results for cold stars with
the APR EoS~\cite{Akmal98}, reference EoS in most papers in the
literature. They are listed in table~\ref{tab:iqcoeff}.

%%%%%%%%%%%%%%%%%%%%%%%%%%%%%%%%%%%%%%%%%%%%%%%%%%%%%%%%%%%%%%%%
\begin{table}
\begin{center}
\begin{tabular}{|c|c|c|c|c|}
\hline
 $a$& $b$  & $c$ &$d$ & $e$  \\ \hline
1.5196& 0.4372& 0.0687& 0.013& 0.000897\\
\hline
\end{tabular}
\caption{Values of the fit parameters in Eq.~(\ref{eq:iqfit}) relating the normalized moment of inertia and quadrupole moment obtained from the results for cold slowly rotating neutron stars with the APR EoS~\cite{Akmal98}.}
\label{tab:iqcoeff}
\end{center}
\end{table}
%%%%%%%%%%%%%%%%%%%%%%%%%%%%%%%%%%%%%%%%%%%%%%%%%%%%%%%%%%%%%%%%%%%%

Considering the difficulty of defining Love numbers for the case of a
rapidly spinning object (see e.g. Pani et al.~\cite{pani15}), we will
focus here on the $\bar{I}$-$\bar{Q}$ relation. Nevertheless, a loss
of universality in this relation would imply a loss of universality in the
more general $\bar{I}$-$\bar{\lambda}$-$\bar{Q}$, too. The results for
different EoSs are shown in Fig.~\ref{fig:iq}.

Results for cold stars are shown in panel (a), for slow and fast
rotating stars, see the symbols for $s_B = 0$. Different colors
represent different EoSs. In addition to the classical nuclear LS and
STOS EoSs, we include other general purpose models, not only purely
nucleonic but, respectively, with $\Lambda$-hyperons and the entire
baryon octet, too, always assuming neutrinoless
$\beta$-equilibrium. The present results, considering in addition
hyperonic EoSs, clearly confirm previous findings that $I$-$Q$
relations are independent of the EoS with frequency dependent fit
coefficients~\cite{Doneva:2013rha}.

In Ref.~\cite{Martinon:2014uua} a study of this relation has been
performed, employing purely nuclear EoSs from Refs.~\cite{Pons1999,
  Pons:2000xf}, this time assuming different realistic entropy per
baryon and electron fraction profiles for the PNS evolution during the
minute following bounce. The main result the authors found was that
universality of the so-called $I$-Love-$Q$ relations is violated in
the early phases of PNS evolution and recovered as soon as the entropy
gradients smoothen out and the star becomes more or less
isentropic. It should then be independent of the exact value of $s_B$.

Our results including hyperonic EoS confirm that indeed, the
$\bar{I}$-$\bar{Q}$-relation for an isentropic star with $s_B = 1$ or
$s_B = 2$ agrees with the result for cold stars. The same is true for
fast rotation, and assuming $\beta$-equilibrium or a constant lepton
fraction $Y_L$ does only induce a small scatter in the results. The
results with constant $s_B = 4$ -- see panels (a) and (b) of
Fig.~\ref{fig:iq} -- although they remain universal in the sense that
there is only a small difference between different EoS, deviate,
however, clearly from the results for cold stars. This can be seen
from Fig.~\ref{fig:iqfit}, too, where we have plotted for the DD2Y
EoS, the relative difference between the numerical results and the fit
function of Eq.~(\ref{eq:iqfit}), $\Delta
\bar{I}/\bar{I}_{\mathit{fit}} = (\bar{I} -
\bar{I}_{\mathit{fit}})/\bar{I}_{\mathit{fit}}$ at different values of
constant entropy per baryon. For $s_B = 1$ or $s_B = 2$, the
deviations remain below 2\%, whereas at $s_B = 4$, they can exceed
10\%.

Both temperature profiles with entropy gradients -- see panels (c) and (d)
-- display obvious deviations from the results for cold stars,
too. With increasing temperature, the differences induced by the
lepton fraction increase, too.

In Ref.~\cite{Martinon:2014uua} the observed deviations from
universality in the early stages of PNS evolution were attributed to
the presence of entropy gradients. Our results suggest a slightly
modified picture, in the sense that universality is not a question of
entropy gradients, but of thermal effects. As we have seen also during
the preceding discussion on maximum masses, at $s_B = 1$ or $s_B = 2$,
which are typical values in the late stages of PNS evolution probed in
Ref.~\cite{Martinon:2014uua}, thermal effects on the EoS and thus on
the star's structure remain small. At higher entropies, thermal
effects start to influence the EoS, thus the star's structure and
universality of $\bar{I}$-$\bar{Q}$ relations are modified. Such
entropy values can be reached in PNSs or merger remnants, depending on many
factors such as the progenitors, rotation or metallicity.

Since the $\bar{I}$-$\bar{Q}$ relation still seems independent of the
employed EoS, it might be tempting to try to obtain another
``universal'' fit, depending this time on rotation frequency and
entropy/temperature. In contrast to the former, neither temperature
nor entropy of the star are quantities which are observationally
accessible. Therefore such a law would not help for data analysis and
we refrain from giving one here. Anyway, in view of the present
results doubts are allowed concerning the relevance of
$\bar{I}$-$\bar{Q}$ relations for analysis of PNS or merger remnant
data, including gravitational wave signals from the last stages of
binary neutron star mergers. Let us stress here that entropy values of
the order of $s_B=4$ are quite realistic in such cases, see e.g. the
simulations in Refs.~\cite{hempel12,steiner13,Perego2014}.

\section{Summary and conclusions}
\label{sec:conclusions}
In this paper we have presented a new consistent general purpose EoS,
including in particular thermal effects. The new EoS, including the
entire baryon octet, is compatible with present constraints from
nuclear physics and neutron star observations.  The complete new EoS
as function of $T,n_B,Y_e$ will be made publicly available in
tabulated form on the \textsc{Compose} database~\cite{Typel2013}, see
appendix~\ref{app:tables} for details.

We have demonstrated the applicability of the new EoS, investigating
maximum masses of hot stars, comparing a purely nuclear EoS with one
including $\Lambda$-hyperons and the new one with all hyperons. To
that end we have applied a numerical code able to provide stationary
models of relativistic rotating stars, including the effect of nonzero
temperature.  The main motivation for studying hot (rotating) stars
are the birth of neutron stars, i.e. the evolution of PNSs, and the
neutron star created in the aftermath of a binary neutron star
merger. In order to correctly identify the configurations which are
secularly stable, we have constructed sequences at different values of
constant total entropy, $S$, in contrast to many previous works
considering constant entropy per baryon, $s_B$.

As we have seen, thermal effects, and a lepton rich environment can
considerably increase the maximally supported mass to a degree
depending on the EoS. The lepton rich environment is important in
particular if hyperons are present. If the entropy per baryon exceeds
roughly $s_B = 2$, thermal effects become important in the EoS, too. Thus
for a total entropy roughly above $5 M_\odot$ thermal effects on the
maximum mass become noticeable. These high temperatures can be reached
in both merger remnants and PNSs depending on the particular
conditions. Let us recall again that previous
works~\cite{Baumgarte1999, Morrison2004, Kastaun2015} suggest that the
main effect stabilizing a merger remnant or a PNS above the maximum
mass of its non-rotating cold neutrinoless $\beta$-equilibrated
counterpart is differential rotation, which we did not consider here.

Following the work by Martinon et al.~\cite{Martinon:2014uua}, the
universality of $I$-$Q$ relations has been tested for fast rotating
hot stars, retrieving their results that a low constant nonzero
entropy does not modify the relations. Universality, tested before
only for purely nuclear models, is maintained in the presence of
hyperons, too. This is, however, no longer true if thermal effects in
the EoS become non negligible, independently of the presence of
entropy gradients, i.e., it also occurs for high, but constant entropies.

\begin{acknowledgments}
We would like to thank Dorota Gondek and Loïc Villain for instructive
discussions. 
This work has been partially funded by the SN2NS project
ANR-10-BLAN-0503, the ``Gravitation et physique fondamentale'' action
of the Observatoire de Paris, and the COST action MP1304 ``NewComsptar''.
\end{acknowledgments}
\appendix
\section{Technical issues of the new EoS table}
\label{app:tables}
The new EoS in its version DD2Y is provided in a tabular form in the
\textsc{Compose} data base, \url{http://compose.obspm.fr} as a
function of $T, n_B, Y_e$. The contribution from electrons is
included. Note that the \textsc{Compose} software allows to calculate additional quantities such as, e.g., sound speed, from those provided in the tables. Please see the \textsc{Compose} manual~\cite{Typel2013} and
the data sheet on the web site for more details about the definition
of the different quantities.

\begin{itemize}
\item The grid is specified as follows:
\begin{table}[h!]
\begin{tabular}{c||c|c|c}
 & $T$ & $n_B$ & $Y_e$ \\ \hline
\# of points & 80 & 302 & 59 \\
Minimum value & 0.1 MeV & $10^{-12} \mathrm{fm}^{-3}$ & 0.01\\
Maximum value & 158.5 MeV & $1.202~\mathrm{fm}^{-3}$ & 0.6\\
Scaling & logarithmic & logarithmic & linear
\end{tabular}
\end{table}
\end{itemize}
\begin{itemize}
\item Thermodynamic quantities provided:
\begin{enumerate}
\item Pressure divided by baryon number density $p/n_B$ [MeV]
\item Entropy per baryon $s/n_B$
\item Scaled baryon chemical potential $\mu_B/m_n - 1$
\item Scaled charge chemical potential $\mu_Q/m_n$
\item Scaled (electron) lepton chemical potential $\mu_L/m_n$
\item Scaled free energy per baryon $f/(n_B m_n) - 1$
\item Scaled energy per baryon $e/(n_B m_n) - 1$
\end{enumerate}
\item Compositional data provided:
\begin{enumerate}
\item Particle fractions of baryons and electrons, $Y_i = n_i/n_B$
\item Particle fractions of deutons ($^2$H), tritons ($ ^3$H), $^3$He, and $\alpha$-particles ($ ^4$He)
\item Fraction of a representative (average) heavy nucleus, together with its average mass number and average charge
\end{enumerate}
Please note that only nonzero particle fractions are listed.
\item Effective Dirac masses $M^*$ of all baryons with nonzero density
  are provided within homogeneous matter.
\end{itemize}
\section{Expressions for the temperature profiles}
\label{app:profiles}
Although they are inspired by results from simulations, for computational simplicity, analytic parameterizations for the temperature profiles, $T_1$ and $T_2$, are employed of the form
\begin{equation}
   T(n_B) = c\, n_B^{\alpha} + d\,n_B + \frac{a \,n_B}{1 + \exp(b\,(n_B-n_0)^2)}~.
\end{equation}
$n_0$ indicates here the saturation density, $n_0 = 0.155
\mathrm{fm}^{-3}$ and the values of the other parameters are listed
below.
\begin{table}[h!]
\begin{tabular}{l|ccccc}
 & $a$ & $b$ & $c$ & $d$ & $\alpha$  \\
 & (MeV fm$^3$) &fm$^6$ &(MeV fm$^{3\alpha}$) & (MeV fm$^3$) &  \\ \hline
$T_1$ &10.01 &26.21 & 77.39&-65.15 & 0.35\\
$T_2$ &470.0 &26.21 & 77.39& -65.15 & $\frac{1}{3}$
\end{tabular}
\end{table}

\bibliography{biblio}

\end{document}